\documentclass{article}

\usepackage{arxiv}

\usepackage{authblk}
\usepackage[utf8]{inputenc} 
\usepackage[T1]{fontenc}    
\usepackage{hyperref}       
\usepackage{url}            
\usepackage{booktabs}       
\usepackage{amsfonts}       
\usepackage{nicefrac}       
\usepackage{microtype}      
\usepackage{lipsum}		
\usepackage{amsmath}
\usepackage{multirow}
\usepackage{physics}
\usepackage[cal=pxtx]{mathalfa}

\usepackage{graphicx}
\graphicspath{{Figures/}}
\usepackage[caption=false]{subfig}
\usepackage[square,numbers]{natbib}
\title{Polynomial Ridge Flowfield Estimation}

\author[1]{A.~Scillitoe\thanks{Corresponding author; Email: \href{mailto:ascillitoe@turing.ac.uk}{ascillitoe@turing.ac.uk}, Web: \href{https://ascillitoe.com}{ascillitoe.com}.}~~}
\author[1,2]{P.~Seshadri}
\author[3]{C.~Y.~Wong}
\author[1,2]{A.~Duncan}
\affil[1]{Data-Centric Engineering, The Alan Turing Institute.}
\affil[2]{Department of Mathematics (Statistics Section), Imperial College London.}
\affil[3]{Department of Engineering, University of Cambridge.}

\usepackage{bm}

  \def\clap#1{\hbox to 0pt{\hss#1\hss}}

\providecommand{\mat}[1]{\bm{#1}}%
\renewcommand{\vec}[1]{\mathbf{#1}}

\providecommand{\mF}{\ensuremath{\mat{F}}}

\providecommand{\mI}{\ensuremath{\mat{I}}}

\providecommand{\mK}{\ensuremath{\mat{K}}}

\providecommand{\mO}{\ensuremath{\mat{O}}}
\providecommand{\mP}{\ensuremath{\mat{P}}}

\providecommand{\mV}{\ensuremath{\mat{V}}}
\providecommand{\mW}{\ensuremath{\mat{W}}}
\providecommand{\mX}{\ensuremath{\mat{X}}}

\providecommand{\vf}{\ensuremath{\vec{f}}}

\providecommand{\vs}{\ensuremath{\vec{s}}}

\providecommand{\vv}{\ensuremath{\vec{v}}}
\providecommand{\vw}{\ensuremath{\vec{w}}}
\providecommand{\vx}{\ensuremath{\vec{x}}}

\begin{document}
\maketitle

\begin{abstract}
Computational fluid dynamics plays a key role in the design process across many industries. Recently, there has been increasing interest in data-driven methods, in order to exploit the large volume of data generated by such computations. This paper introduces the idea of using spatially correlated polynomial ridge functions for rapid flowfield estimation. Dimension reducing ridge functions are obtained for numerous points within training flowfields. The functions can then be used to predict flow variables for new, previously unseen, flowfields. Their dimension reducing nature alleviates the problems associated with visualising high dimensional datasets, enabling improved understanding of design spaces and potentially providing valuable physical insights.

The proposed framework is computationally efficient; consisting of either readily parallelisable tasks, or linear algebra operations. To further reduce the computational cost, ridge functions need only be computed at only a small number of subsampled locations. The flow physics encoded within covariance matrices obtained from the training flowfields can then be used to predict flow quantities, conditional upon those predicted by the ridge functions at the sampled points.

To demonstrate the efficacy of the framework, the incompressible flow around an ensemble of aerofoils is used as a test case. On unseen aerofoils the ridge functions' predictive accuracy is found to be reasonably competitive with a state-of-the-art convolutional neural network (CNN). The local ridge functions can also be reused to obtain surrogate models for integral quantities such a loss coefficient, which is advantageous in situations where long-term storage of the CFD data is problematic. Finally, use of the ridge framework with varying boundary conditions is demonstrated on a three dimensional transonic wing flow.
\end{abstract}


\section{Introduction}
\label{sec:intro}

The industrial penetration of Reynolds Averaged Navier-Stokes (RANS) has gone far beyond its mainstay of aerospace engineering. Today it is being heavily used in automative design \cite{Kobayashi2009}, building design \cite{Ding2019}, ventilation system design \cite{Ramponi2012}, marine propeller design \cite{Saha2019} and more recently in the understanding of oral transmissions \cite{Dbouk2020,Dbouk2020b}. In using a turbulence model, RANS avoids the need to resolve pertinent length- and time-scales of turbulence; significantly slashing the cost of generating a flowfield. That said, owing to the size and increasingly realistic mesh topologies, these RANS computations can themselves take several hours on modern clusters. Additionally, in both design and analysis contexts, there is a need to assess multiple geometry and boundary conditions, adding to the computational cost. 

It is therefore unsurprising that many have sought to develop ways of (i) efficiently representing a flowfield and (ii) approximating it for a new set of boundary conditions or geometry definition. Falling under the remit of sparse reconstruction, there exists a variety of approaches for reconstructing full unsteady flowfield information from a limited subset of time-dependent measurements. A suitable basis must first be chosen, onto which the sparse measurements are projected in order to obtain a full reconstruction. Often the chosen basis is linear, and involves a direct computation of the singular value decomposition. For example; \citet{Manohar2018} use proper orthogonal decomposition to reconstruct full vorticity fields of laminar flow over a cylinder, while \citet{Discetti2018} reconstruct turbulent velocity fields from sparse sensors. More recently, supervised machine learning methods have been used to provide a non-linear framework, with \citet{Erichson2020} using a shallow neural network to reconstruct the laminar flow over a cylinder. 

Meanwhile, in a range of fields, including uncertainty quantification, design optimisation, and sensitivity analysis, lower fidelity surrogate models (or emulators) are constructed from existing full-field simulation data. Deep neural network architectures, routinely used in data mining, have been used with considerable success as a function approximation technique for high-dimensional physics derived datasets \cite{Raissi2018,Agostini2020}. To increase accuracy, a network can be extended to arbitrary depth and width, however this leads to many weights and a very large amount of training data being required to alleviate overfitting. Recently, \emph{convolutional neural networks} (CNN's) have been shown to be a promising approach to ameliorate this problem for CFD applications. CNN's use convolutional layers to take advantage of local spatial coherence in the input. These layers, combined with successive spatial resizing of the input data, can significantly reduce the number of weights compared to fully connected neural networks.  \citet{Guo2016} introduced the idea of using a CNN to learn a mapping between an object's geometric representation and the flowfield around it. \citet{Bhatnagar2019} and \citet{Thuerey2020} recently built upon this work by introducing boundary conditions as an additional input. Such approaches offer accurate and fast data-driven flowfield predictions, allowing for near-immediate feedback for real-time design iterations. Following a different strategy, \citet{Tompson2017} replace the pressure projection step in a CFD solver with a CNN in order to accelerate simulations, whilst \citet{Sekar2019} use a CNN for the inverse design of an aerofoil, and \citet{Jin2018} use a CNN to estimate velocity fields from surface pressure distributions.

Although applications of deep learning networks to real-world problems have become ubiquitous, our understanding of why they are so effective is lacking\cite{Sejnowski2020}. Techniques for interpretation are available \cite{Zhang2018}, but they may not be particularly accessible for those without some specialised deep learning knowledge. Moreover, for problems with large numbers of features, widely adopted approaches such as those based on Shapley values \cite{Lundberg2017} quickly become intractable. Consequently, in many fields of engineering, there is still a demand for more readily interpretable surrogate models, even if these bring a slight trade-off in predictive accuracy. One approach to easing the challenge of interpretation, is to reduce the dimensionality of the problem. By reducing the dimensionality of the input space, data-driven ridge functions \cite{pinkus2015ridge} facilitate straightforward visualisation and understanding, whilst at the same time reducing data requirements. \citet{delRosario2017} show how ridge functions can offer important insights during the conceptual design phase, and in \cite{scillitoe2020design}, ridge functions aid exploration of a stagnation temperature probe's design space. \citet{wong2020embedded} introduce the idea of using dimension reduction to estimate flow quantities, embedding polynomial ridge functions on an aerofoil surface to estimate the pressure distribution. In this paper, we develop spatially correlated ridge functions, with the target output being flowfield variables such as pressure, velocity or even turbulent variables. 

To demonstrate the utility of the proposed framework, we begin by exploring the prediction of flowfields around a set of aerofoils. The design of efficient aerofoils is important since their use is ubiquitous across many industries; for example aircraft wings and controls surfaces, car spoilers, and the gas turbine blades found in aircraft engines and power generation turbines. To provide a point of reference, we compare and contrast the proposed approach to a state-of-the-art flowfield prediction framework based on a convolutional neural network. In addition to examining the predictive accuracy, we explore how the dimension reducing nature of the learned model aids understanding of the existing training dataset. To assess the limitations of the proposed approach, we then tackle the transonic flow over a three dimensional wing, with the inflow boundary conditions varied. We examine how the underlying polynomials perform when extrapolating away from the training data, and discuss approaches to handling this situation. For a hands-on exploration of the framework, example code and an interactive web app are available at \url{github.com/ascillitoe/flowfield_approx}.

\section{Mathematical foundations}
\label{sec:method}
In this section, we detail some of the mathematical ideas that underpin this paper. 

\subsection{Flowfield representation}
It will be useful to represent a steady flowfield as a set of related scalar-field quantities. These include, but are not limited to, static pressure $p$, density $\rho$, a velocity component i.e. $v_x$, and even the turbulent viscosity $\nu_t$. These fluid properties are assumed known for any particular discretised location $\vs \in \mathbb{R}^{3}$ within the three-dimensional flow domain $\mathcal{S}$. We consider $\vs$ to be synonymous with the number of nodes in the flowfield, where $\vs_{1}$ corresponds to the first coordinate and $\vs_{N}$ corresponds to the last.

Ignoring model parameters and limiters embedded within any modern flow-solver, one can group the parameters influencing a flowfield into (i) those that govern the mesh and its associated geometry, and (ii) those that set the boundary conditions. For simplicity, we condense the former into a vector of shape parameters $\vx \in \mathbb{R}^{d}$ arising from an appropriately chosen design space $\mathcal{X}$, e.g., Hicks-Henne bump functions, free-form deformation amplitude, or the inputs to an autoencoder that characterises the geometry \cite{wang2021airfoil}. We characterise the boundary conditions by a vector $\vv \in \mathbb{R}^{b}$ where $b$ represents the number of input boundary conditions, e.g., inlet stagnation pressure, exit static pressure, isothermal walls, etc.

Then, following the statement at the start of this section, we can represent the steady flowfield as a set of related vectors
\begin{equation}
\mathcal{F}=\left\{ \;  \left[\begin{array}{c}
p\left(\vx,\vv; \vs_{1}\right)\\
\vdots\\
p\left(\vx,\vv ; \vs_{N}\right)
\end{array}\right],\left[\begin{array}{c}
\rho\left(\vx, \vv, \vs_{1}\right)\\
\vdots\\
\rho\left(\vx, \vv ;\vs_{N}\right)
\end{array}\right],\left[\begin{array}{c}
v_{x}\left(\vx,  \vv ; \vs_{1}\right)\\
\vdots\\
v_{x}\left(\vx, \vv ;\vs_{N}\right)
\end{array}\right], \; \ldots \;,  \left[\begin{array}{c}
\nu_t \left(\vx,\boldsymbol{\zeta};s_{1}\right)\\
\vdots\\
\nu_t \left(\vx,\boldsymbol{\zeta}; s_{N}\right)
\end{array}\right] \; \right\}.
\end{equation}
For generality, we refer to any scalar field quantity by $f$ and thus each of the vectors in $\mathcal{F}$ is generalized via
\begin{equation}\label{equ:vector}
\vf = \left[\begin{array}{c}
f \left(\vx, \vv;\vs_{1}\right)\\
\vdots\\
f \left(\vx,\vv; \vs_{N}\right)
\end{array}\right].
\end{equation}
Each element in \eqref{equ:vector} is a scalar-valued function depending only on $\vx$ and $\vv$. Identifying a suitable surrogate model for $\vf$ is challenging because it effectively has $\mathbb{R}^d \times \mathbb{R}^b \times \mathbb{R}^{N}$ degrees of freedom, leading to an insuperable number of model evaluations. In this paper, we demonstrate how subspace-based dimension reduction approaches can be used to thwart the prohibitive cost associated with constructing such a surrogate. In what follows, we present a class of techniques for approximating each element. Without loss in generality, we assume the boundary conditions are fixed, such that each scaler-valued function depends only on $\vx$. However, in Section~\ref{sec:results2} we shall consider an alternative case where $\vv$ is varied whilst $\vx$ is fixed. 

\subsection{Ridge approximations}
\label{sub:ridges}

Central to this paper is the notion that a scalar-valued high-dimensional function, such as the static pressure at a given computational node $\vs_{i}$, can be approximated by a low-dimensional function defined over a sub-manifold
\begin{align}
f \left( \vx; \vs_{i} \right) & \approx p_{i} \left( \mW_{i}^{T} \vx ; \vs_{i} \right),
\label{equ:ridge_approx}
\end{align}
where $i = \left( 1, \ldots, N \right)$. Here $\mW_{i} \in \mathbb{R}^{d \times n}$ is an orthonormal matrix where $n \ll d$, and $p_i
$ is a function in $\mathbb{R}^{n}$. We define the approximation in \eqref{equ:ridge_approx} as a ridge approximation, an approximate form of what Pinkus \cite{pinkus2015ridge} terms a generalised ridge function. The intuitive computational advantage behind ridge approximations is that instead of estimating a function in $\mathbb{R}^d$, we approximate it in $\mathbb{R}^n$, which is intrinsically more data efficient. Numerous approaches are available in literature for identifying such structure using only data-driven techniques. These require input-output data pairs of the form $\left\{\vx_m, f\left( \vx_m; \vs_{i}\right)\right\}_{m}^{M}$, where $M$ represents a suitable number of design of experiments required to approximate the function in $\mathbb{R}^n$. Approaches rooted in classical regression include sliced inverse regression \cite{li1991sliced}, principal Hessian directions \cite{li1992principal}, contour regression \cite{li2005contour}, sliced average variance estimation \cite{cook1991sliced}, and minimum average variance estimation \cite{xia2002adaptive}. More recent approaches include the works of \citet{constantine2017near} and \citet{hokanson2018data} that assume $p_{k}$ is a polynomial, and explore optimisation techniques over the Grassmann manifold to identify both the coefficients of the polynomial and the dimension reducing subspace $\mW$. Algorithms for identifying such ridge structure exist even for Gaussian process models (see \cite{seshadri2019dimension} and \cite{liu2017dimension}). 

\begin{figure}[ht]
	\centering
     \includegraphics[width=0.55\linewidth]{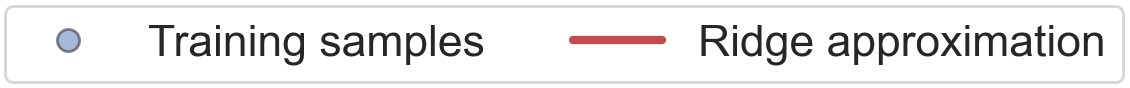} \vspace{-5pt} \\
     \subfloat[Equal contributions from all input dimensions, with a unit vector $\mW=\vec{1}/\sqrt{25}$
  		\label{subfig:blade_ridge_a}]{
  		\includegraphics[width=0.48\linewidth]{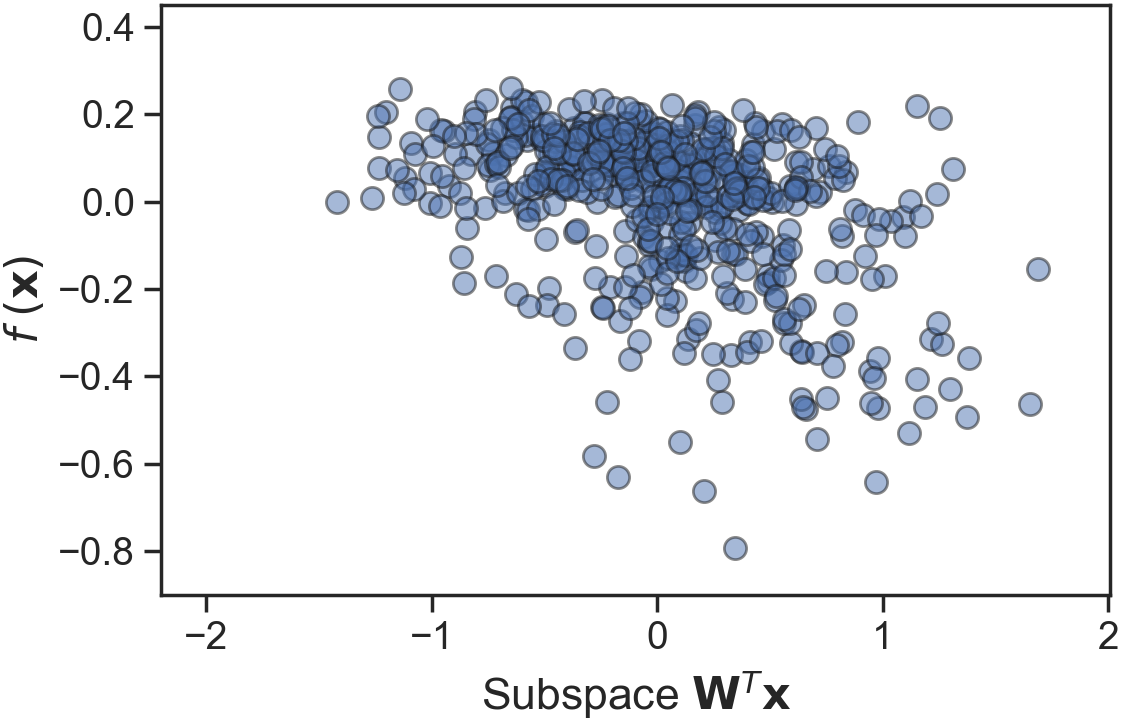}}   \hfil 
     \subfloat[Dimension reduction to obtain $\mW$, with a cubic orthogonal polynomial fitted over the subspace
  		\label{subfig:blade_ridge_b}]{%
  		\includegraphics[width=0.48\linewidth]{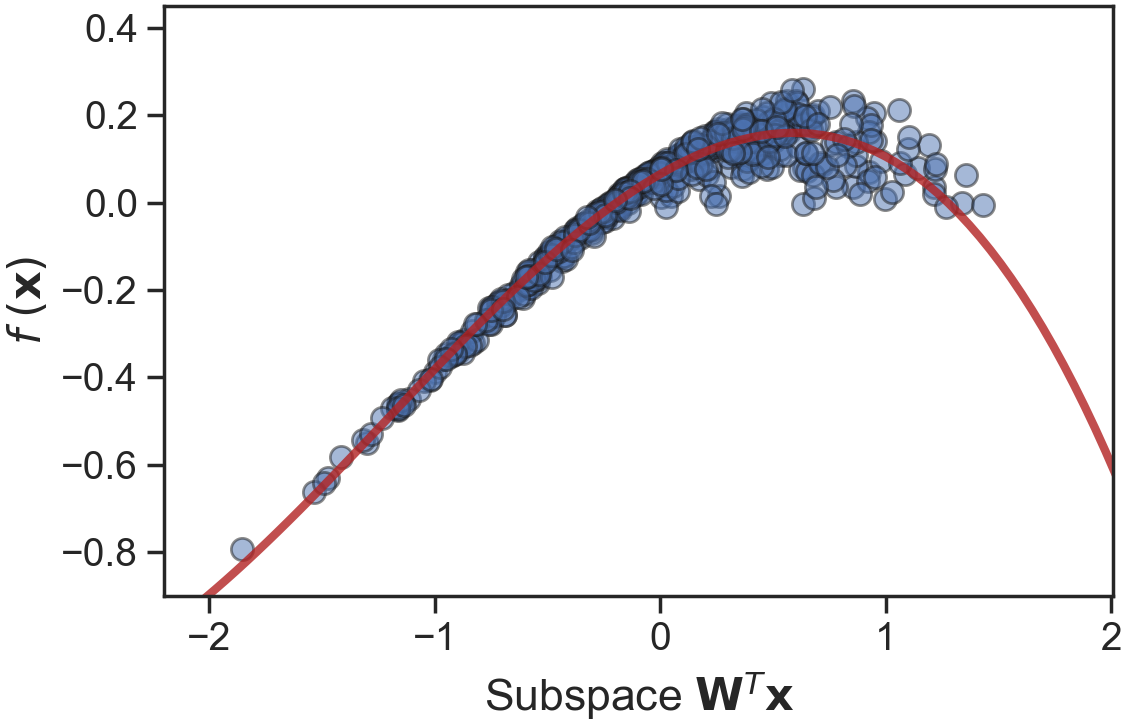}}
 
	\caption{Subspace-based dimension reduction applied to a turbo-machinery problem\cite{seshadri2018turbomachinery}. The output scalar-valued quantity of interest $f(\vx)$, the efficiency of a turbo-machinery blade, is a function of the $d=25$ design parameters in $\vx \in \mathbb{R}^{25}$. The quantity $f$ is projected over two different subspaces $\mW^T\vx$.	In a), $\mW$ is chosen so that the subspace is formed from an equal contribution of all 25 parameters, whilst in b), a single linear combination of the $25$ parameters which best describes the variation in $f$ is obtained.}
	\label{fig:blade_ridge}
\end{figure}

To provide an intuitive understanding of what these methods can offer, in Figure~\ref{fig:blade_ridge} we apply subspace-based dimension reduction to a dataset obtained from \cite{seshadri2018turbomachinery} and publicly available in \url{github.com/equadratures/data-sets}. When a suitable subspace $\mW$ is found, the $M$ number of input-output training data pairs $\left\{\vx_m, f_m \right\}_{m=1}^{M}$ collapse onto a ridge over the subspace, where they can be used to obtain a functional approximation (a polynomial in this case). Plots of function values over this subspace are called \emph{sufficient summary plots}\cite{constantine2017near}, and are useful in visualising the behaviour of functions in high dimensions.

\subsection{Polynomial ridge approximations}
\label{sub:poly_ridge}
Let us assume access to a training dataset $\left\{ \mX, \mF \right\}$, where
\begin{equation}
\mX=\left[\begin{array}{ccc}
| &  & |\\
\vx_{1} & \ldots & \vx_{M}\\
| &  & |
\end{array}\right], \; \; \; \; \;  \; \; \mF=\left[\begin{array}{ccc}
- & \vf_{1}^{T} & -\\
 & \vdots\\
- & \vf_{N}^{T} & -
\end{array}\right] \; = \;\left[\begin{array}{ccc}
| &  & |\\
\tilde{\vf}_{1} & \ldots & \tilde{\vf}_{M}\\
| &  & |
\end{array}\right]
\end{equation}
with $\mX \in \mathbb{R}^{d \times M}$ and $\mF \in \mathbb{R}^{N \times M}$. As before $N$ represents the number of spatial nodes; $d$ the dimension associated with the design space $\vx \in \mathcal{X}$, and $M$ the number of distinct flow-fields, each corresponding to a new design. At an isolated node $s_i$ across all the flow-fields, we wish to identify a polynomial ridge approximation of the form given in \eqref{equ:ridge_approx}. To elaborate on the form of this ridge approximation, we rewrite it as
\begin{equation}
f \left( \vx; \vs_{i} \right) \approx \mP\left( \mW_{i}^{T} \vx^{\ast} \right) \boldsymbol{\alpha}_{i}
\label{equ:ridge_vp}
\end{equation}
where 
\begin{equation}
\mP\left(\mW_{i}^{T} \mX \right) = \left[\begin{array}{ccc}
\boldsymbol{\phi}_{1}\left( \mW_{i}^{T}\vx_{1}\right) & \ldots & \boldsymbol{\phi}_{L}\left(\mW_{i}^{T}\vx_{1}\right)\\
\vdots & \vdots & \vdots\\
\boldsymbol{\phi}_{1}\left(\mW_{i}^{T} \vx_{M}\right) & \ldots & \boldsymbol{\phi}_{L}\left(\mW_{i}^{T} \vx_{M}\right)
\end{array}\right]
\end{equation}
represents a Vandermonde-type matrix with $n$-variate polynomial basis terms $\left\{\boldsymbol{\phi}_{1}, \ldots, \boldsymbol{\phi}_{L} \right\}$ with a cardinality of $L$, with unknown polynomial coefficients $\boldsymbol{\alpha}_{i} \in \mathbb{R}^{L}$. The cardinality here is defined as the total number of basis terms, and will vary depending on the highest degree in along each dimension and the nature of the interactive polynomial terms between different dimensions. For further details on the construction of multivariate polynomials, we refer the interest reader to Section 1.1 in \cite{seshadri2017effectively}. The intention behind phrasing our objective as \eqref{equ:nlinlsq}, is that for a new design $\vx^{\ast}$, we can estimate the scalar field value at any node $s_i$ using the computed polynomial ridge approximation. To ensure solutions to this problem are feasible, it will be useful to ensure that the polynomial matrix $\mP$ is well conditioned. This can be aided by adopting orthogonal polynomial basis terms---i.e., Legendre, Hermite or Jacobi---as these have demonstrably lower condition numbers than their monomial counterparts. In this paper, we utilise Legendre polynomials. This has the additional advantage of being able to rapidly yield moments and sensitivities (see \cite{wong2021extremum}).

The ridge approximation in \eqref{equ:ridge_vp} can be obtained by solving the non-linear least squares problem 
\begin{equation}
\underset{\mW_{i}\in\mathbb{R}^{d\times n},\;\bm{\alpha}_{i} \in \mathbb{R}^{L}}{\textrm{minimise}} \; \left\Vert \vf_{i} - \mP\left(\mW_{i}^{T} \mX\right) \boldsymbol{\alpha}_{i} \right\Vert _{2}^{2},
\label{equ:nlinlsq}
\end{equation}
which can be simplified to an optimisation problem over the Grassmann manifold via the variable projection method\cite{golub2003separable}  
\begin{align}
& \underset{\mW_{i}\in\mathbb{R}^{d\times n} }{\textrm{minimise}} \; \left\Vert \vf_{i} - \mP\left(\mW_{i}^{T} \mX\right) \mP\left(\mW_{i}^{T} \mX\right)^{\dagger} \vf_{i} \right\Vert _{2}^{2} \\
\Rightarrow \; & \underset{\mW_{i}\in\mathbb{R}^{d\times n} }{\textrm{minimise}} \; \left\Vert \left( \mI - \mP\left(\mW_{i}^{T} \mX\right) \mP\left(\mW_{i}^{T} \mX\right)^{\dagger}\right) \vf_{i} \right\Vert _{2}^{2} \\
\Rightarrow \; & \underset{\mW_{i}\in\mathbb{R}^{d\times n} }{\textrm{minimise}} \; \left\Vert  \mO^{\perp}_{\mP \left(\mW_{i}^{T} \mX\right) }  \vf_{i} \right\Vert _{2}^{2}
\label{equ:nlinlsq_sep}
\end{align}
where the superscript $\dagger$ denotes the matrix pseudoinverse, and $\mO^{\perp}$ is the orthogonal projector onto the complement of the column space of $\mP$. Gradients for this objective can be readily computed and used in a Gauss-Newton algorithm over the Grassman manifold, as demonstrated in  \cite{hokanson2018data}. The computation of the coefficients $\boldsymbol{\alpha}_{i}$ trivially follows. This enables us to compute the scalar field quantity at a given node $i$ for a new design $\vx^{\ast}$ with $\mP\left( \mW_{i}^{T} \vx^{\ast} \right) \boldsymbol{\alpha}_{i} $ and by extension the scalar field across all $N$ nodes via
\begin{equation}
\left[\begin{array}{c}
f \left(\vx^{\ast}; \vs_{1}\right)\\
\vdots\\
f \left(\vx^{\ast}; \vs_{N}\right)
\end{array}\right] \approx \left[\begin{array}{c}
\mP\left(\mW_{1}^{T}\vx^{*}\right)\boldsymbol{\alpha}_{1}\\
\vdots\\
\mP\left(\mW_{N}^{T}\vx^{*}\right)\boldsymbol{\alpha}_{N}
\end{array}\right],
\end{equation}
which by construction constraints all subspaces $\mW_1, \ldots, \mW_{N}$ to have the same dimension. While the calculation of polynomial ridge approximations across all $N$ nodes is an embarrassingly parallel operation---as each the input-output data pairs at each node are treated independently---it will be useful to leverage the spatial correlations in a given scalar field to reduce the number of times \eqref{equ:nlinlsq_sep} has to be solved.

\subsection{Spatial correlations}
\label{sub:covar}
Subspace-based dimension reduction approaches and the related active subspaces \cite{constantine2015active} have enjoyed tremendous computational success over the past few years---ushering in a shift from prior sparse grids and full-space design of experiment approaches. As a general observation, within computational fluid dynamics applications, these methods have been primarily applied on integral quantities of scalar field outputs, i.e., efficiency, lift, drag, etc. \citet{wong2020embedded} recognised this, and posited that approximating the composite scalar fields of such integral quantities using dimension reducing subspaces may not only offer greater insight into significant flow features, but may even further reduce the number of simulations required to do so. One important idea born from their work is that the ridge approximations of nodes that are adjacent to each other in a flowfield are likely to be similar in the way they depend on input parameters. In locally smooth regions of the flowfield, nearby nodes are predominantly affected by a similar subset of inputs, resulting in similarity in their ridge subspaces. This notion is quantified via the subspace distance, and in \citet{wong2020embedded} the authors demonstrate that this property enables efficient compression and recovery of flow fields. In what follows, we build upon this idea using a different formalism, leveraging some of our preliminary work in \cite{scillitoe2021instantaneous}.

Using the data available in $\mF$, we construct a sample covariance matrix $\mK \in\mathbb{R}^{N \times N}$ of the form
\begin{equation}
\mK = \frac{1}{M} \sum_{m=1}^{M} \left( \tilde{\vf}_{m} - \mathbb{E}\left[ \tilde{\vf}_{m} \right] \right) \left( \tilde{\vf}_{m} - \mathbb{E}\left[ \tilde{\vf}_{m} \right] \right)^{T},
\label{equ:cov_def}
\end{equation}
where $ \mathbb{E}\left[\tilde{\vf}_{m}\right]$ denotes the mean of the $m$-th scalar field across all the $N$ nodes. It is important to note that this covariance matrix captures the spatial correlations across all the nodes for a given scalar field quantity. Its diagonal represents the sample variances in the scalar field at each of the $N$ nodes. 

We partition the scalar field across all $N$ nodes into two subsets: a smaller subset $\mathcal{J}$ with $J$ nodes $\left(\hat{\vs}_{1}, \ldots, \hat{\vs}_{J}\right)$, and a subset $\mathcal{R}$ containing the remaining $N-J$ nodes $\left( \vs'_{1}, \ldots, \vs'_{N-J}\right)$ with $J<<N$. We assume that the scalar field for any design $\vx^{\ast}$ across all $N$ nodes can be expressed as a sample from the multivariate normal distribution $\mathcal{N} \left( \mathbb{E}\left[ \vf \right], \mK \right)$, where the first argument represents the mean and the second the covariance. The covariance matrix in \eqref{equ:cov_def} can be partitioned into $(N-J) \times (N-J)$ and $J \times J$ blocks
\begin{equation}\label{eqn:covar_blocks}
\mK = \left[\begin{array}{cc}
\mK_{11} & \mK_{12}\\
\mK_{12}^{T} & \mK_{22}
\end{array}\right], \; \; \; \; \textrm{where} \; \; \; \mK_{11}\in \mathbb{R}^{(N-J)\times (N-J)}\; \; \textrm{and} \; \; \mK_{22} \in \mathbb{R}^{J \times J}.
\end{equation}
As the scalar fields arising from this partition will have Gaussian marginals, we can use the Schur complement to approximate the scalar field at the nodes in $\mathcal{R}$ by computing the ridge approximations at the nodes in the smaller set $\mathcal{J}$ only
\begin{equation}\label{eqn:schur}
\left[\begin{array}{c}
f\left(\vx^{*}; \vs'_{1}\right)\\
\vdots\\
f\left(\vx^{*}; \vs'_{N-J}\right)
\end{array}\right]\approx \mK_{12} \mK_{22}^{-1}\left(\left[\begin{array}{c}
\mP\left(\mW_{1}^{T} \vx^{*};\hat{\vs}_{1}\right)\boldsymbol{\alpha}_{1}\\
\vdots\\
\mP\left(\mW_{J}^{T} \vx^{*};\hat{\vs}_{J}\right) \boldsymbol{\alpha}_{J}
\end{array}\right]\right).
\end{equation}
In a nutshell, this implies that we only need to solve \eqref{equ:nlinlsq_sep} $J$ times rather than that $N$. However, this approach relies on the assumption that a suitable $\mW_i$ is found at each point in $\mathcal{J}$, such that the projection $\mW_i^T\vx$ at each point in the subset explains a large percentage of the total variance at that point. Without this assumption, the resulting correlation matrix $\mK$ will become increasingly diagonal in higher dimensions, and we will be unable to recover the scalar field at the remaining nodes in $\mathcal{R}$. To alleviate this issue, poor quality ridge approximations are removed from the smaller subset $\mathcal{J}$, and added back into $\mathcal{R}$. Various heuristics can be used to assess the quality of ridge approximations. In this paper, we use the training $R^2$ scores, as discussed in Section~\ref{sub:spatial_covar}. Finally, for the stability of this procedure it is important that the inverse of $\mK_{22}$ is computed carefully. In the present work, this is done via Cholesky decomposition, with a small constant added to the diagonal to prevent singular matrices.

\section{Computational setup}
\label{sub:data}

In this section we outline the two test cases examined in this paper, in addition to summarising the implementation of the ridge approximation framework and the CNN it will be compared to.

\subsection{Test Cases}

Within an engineering context, flowfield predictions around a given object are often required at a range of operating conditions. Whereas in other cases, flowfield predictions may be required at a given operating condition, whilst the object itself is altered. To examine both scenarios we use two test cases; 1) a two dimensional subsonic aerofoil flow, with a large number of different aerofoil designs; and 2) a three dimensional transonic wing flow, with the freestream conditions varied.  

\subsubsection{Exploring the design of a subsonic aerofoil}

We start with the well known NACA0012 aerofoil, discretised with a 449x129 curvilinear C-grid\footnote{This grid is used as a verification case by the AIAA Fluid Dynamics Technical Committee Turbulence Model Benchmarking Working Group (TMBWG) and is available from \url{turbmodels.larc.nasa.gov/naca0012_grids.html}. \citet{Diskin2016} show lift and drag coefficients to be sufficiently grid independent at the 449x129 grid resolution.}, and deform it to obtain an ensemble of aerofoil designs. The aerofoil surface is deformed using $d=50$ Hicks-Henne bump functions \cite{Hicks1978}
\begin{equation}
s_2(s_1) = s_{2,base}(s_1) + \sum_{j=1}^d \beta_j \mathit{b}_j(s_1),
\end{equation}
where $s_{2,base}$ are the $s_2$ coordinates of the baseline aerofoil, $\mathit{b}_j$ is the $j^{th}$ bump function, and $s_1$ and $s_2$ are normalised by the aerofoil's axial chord length $C_1$. The bump amplitudes $[\beta_1,\dots,\beta_d]$ are then stored within the input vector $\mathbf{x}_m\in \mathbb{R}^d$ for each $m^{th}$ design.

\begin{figure}[ht]
\centering
\includegraphics[width=.85\linewidth]{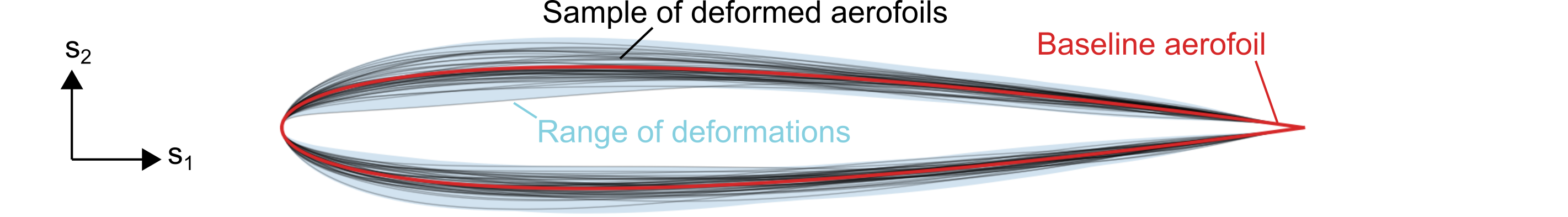}
\caption{Deformations made to the NACA0012 aerofoil. Fifty randomly selected deformed designs are shown.}
\label{fig:aerofoil_deforms}
\end{figure}

To generate a dataset for training and testing, we create a ($M = 2000$)-point design of experiments (DoE) with uniformly distributed Monte Carlo samples for $\mathbf{x}$. A random sample of the resulting aerofoil designs, as well as the full range of deformations, is shown in Figure~\ref{fig:aerofoil_deforms}. Flowfields are simulated for each design using the incompressible solver of the SU2 CFD code \cite{Economon2016}. The commonly used Spalart–Allmaras RANS model\cite{Spalart1992} is used to represent the effects of turbulence, with the freestream turbulence viscosity ratio set to $(\nu_t/\nu)_\infty=5$. The freestream velocity magnitude is set to $U_\infty=1$ m/s, and the reference static pressure is set to zero. The laminar viscosity $\nu_\infty$ is then set to give a Reynolds number of $Re=U_\infty C_1/\nu_\infty=6\times10^6$. Each design is run at three angles of incidence $\alpha_\infty=0^{\circ}$, $10^{\circ}$ and $15^{\circ}$, leading to a dataset consisting of 6000 flowfields in total.

\begin{figure}[ht]
\centering
\includegraphics[width=.65\linewidth]{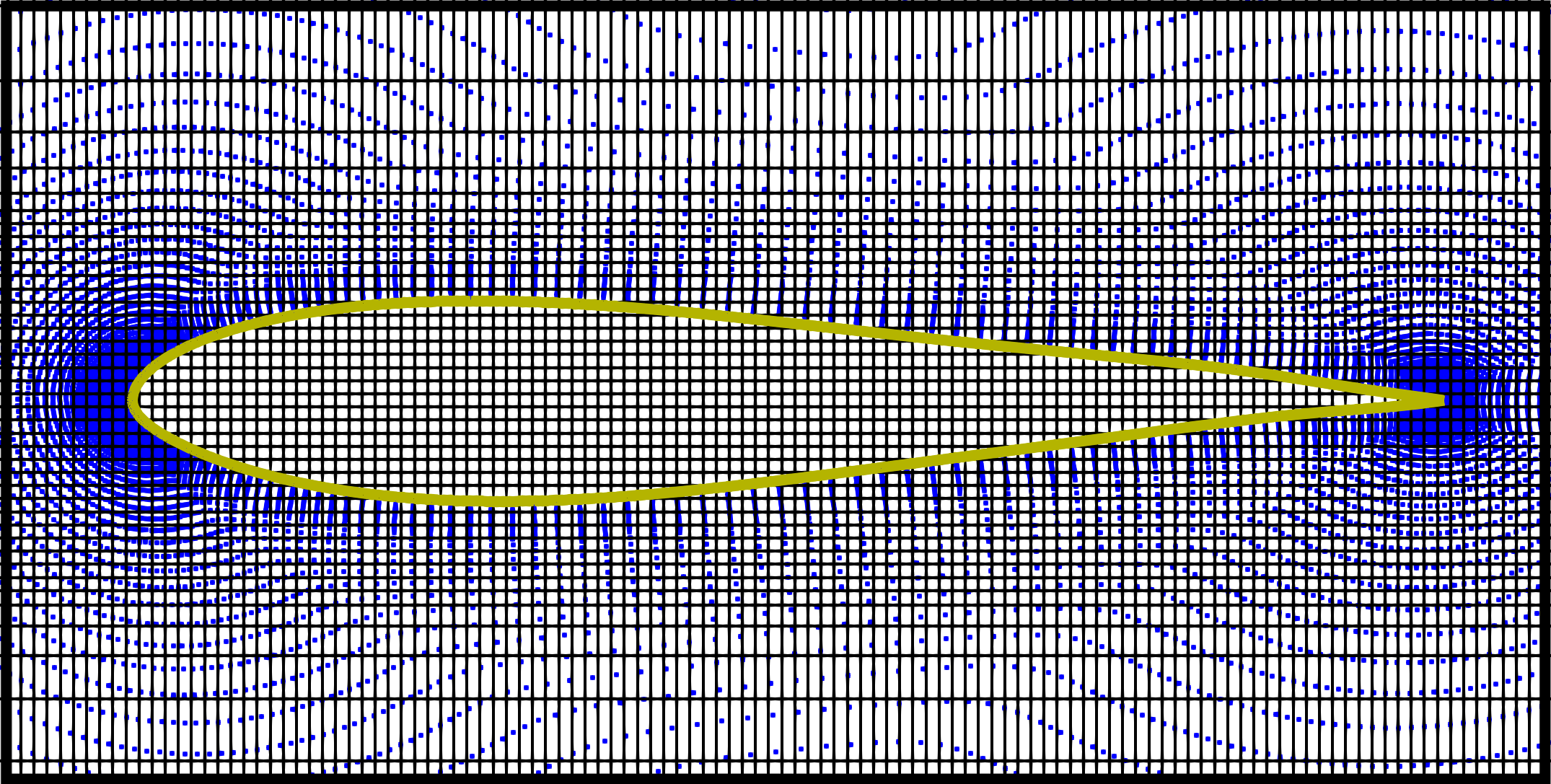}
\caption{Zoomed-in view of the discrete curvilinear C-grid representation of a deformed aerofoil (blue points), and the Cartesian grid (black) the flow variables are resampled onto. The aerofoil boundary is shown in yellow.}
\label{fig:meshes}
\end{figure}

The grid deformations performed when perturbing the baseline aerofoil to reach each new design mean that, even away from the aerofoil surface, grid points are at a slightly different location for each design. Since we wish to learn a functional mapping for the flowfield variables at fixed points in space, we re-sample\footnote{Resampling is performed with the pyvista python library \cite{Sullivan2019}, which uses linear interpolation for resampling.} each flowfield onto the 90x318 Cartesian grid shown in Figure~\ref{fig:meshes}, and remove points lying inside the solid region of the aerofoil. We choose a Cartesian grid here since we are interested in full flowfield visualisations, but the approach is equally applicable to individual points, surfaces (see Ref. \cite{wong2020embedded}) or planes.


After the above procedure we are left with $N=90\times318=28.6\times10^3$ output values $f_{i,m}$ for each input $\mathbf{x}_m$, forming $M$ input/output pairs $(\mathbf{x}_m,\mathbf{f}_m)_{m=1}^M$, for each of the three angles of incidence investigated. The scalar $f_{i,m}$ is the field variable at the $i^{th}$ grid point for the $m^{th}$ design, where we take the static pressure $p$, velocities $u,v$, and turbulent viscosity $\nu_t$ as the field variables to predict. We normalise the aforementioned variables by taking the static pressure coefficient $C_p = (p-p_\infty)/(p_{0_\infty} - p_\infty)$, axial velocity ratios $u/U_\infty$, $v/U_\infty$, and turbulent viscosity ratio $\nu_t/\nu_\infty$. The $M=2000$ designs are further split up into $M_{train}=500$ training designs and $M_{test}=1500$ testing designs.

\subsubsection{Varying freestream conditions for a transonic wing}
\label{sub:onera}

\begin{figure}[t] 
\centering
\includegraphics[width=\linewidth]{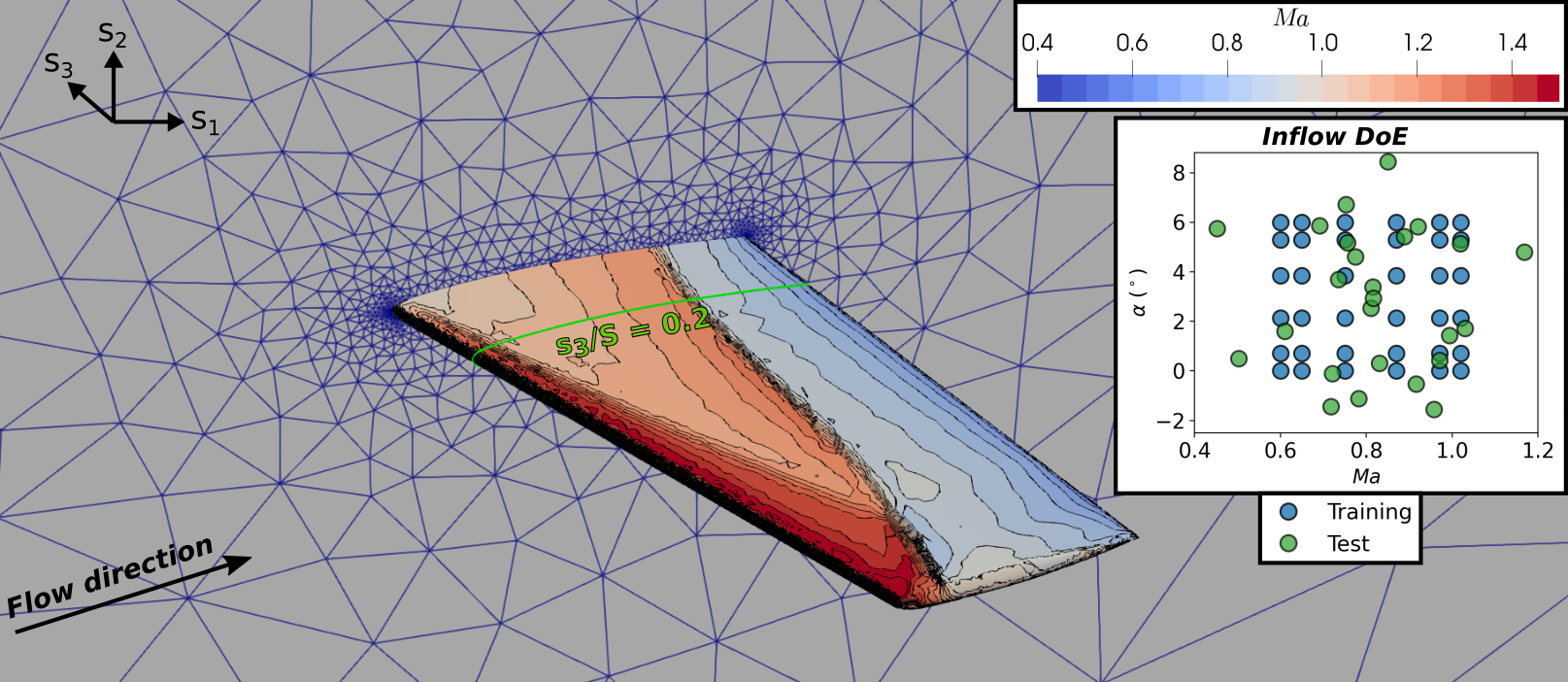}
\caption{Mach number contours on the suction surface of the ONERA M6 transonic wing, under the baseline freestream conditions of $Ma_{\infty}=0.8395$ and $\alpha_{\infty}=3.06^{\circ}$. The unstructured grid is shown on the $s_3=0$ symmetry plane, and the $s_3/S=0.2$ slice is highlighted. The design of experiments (DoE) used to generate the different freestream conditions for training and test data is also shown in the inset figure.}\label{fig:onera_viz}
\end{figure}

To explore predictions with varying operating conditions, we take the compressible inviscid flow around the well known ONERA M6 transonic wing test case \cite{Schmitt1979}. The salient details of this case are presented in Figure~\ref{fig:onera_viz}. It is discretised by a $N=108.4\times10^3$ point unstructured grid, with no boundary layer grid necessary due to the inviscid nature of the flow. Freestream boundaries are placed $10.5$ root chord length's away from the wing, with the freestream Mach number and angle of incidence set at $Ma_{\infty}=0.8395$ and $\alpha_{\infty}=3.06^{\circ}$ for the baseline case. A span-wise symmetry boundary is then placed at $s_3=0$. The flows are simulated using the compressible inviscid solver of the SU2 CFD code\cite{Economon2016}, and in Figure~\ref{fig:onera_viz} the resulting Mach contours are shown on the surface of the wing for the baseline case. The flowfield here bears a close resemblance to the real viscous flow, with a characteristic ``lambda shock'' pattern visible on the upper surface of the wing.

Shown in the inset figure in Figure~\ref{fig:onera_viz} is the design of experiments used to explore the freestream conditions. For the training data, operating points for $Ma_{\infty}$ and $\alpha_{\infty}$ are obtained by applying six-point Gauss-Lobatto quadrature rules over the intervals $Ma_{\infty} \in [0.6,1.02]$ and $\alpha_{\infty} \in [0,6]$, leading to a total of $M_{train}=36$ training flows. For testing, $M_{test}=25$ points are sampled from the multivariate normal distribution $\mathcal{N}(\boldsymbol{\mu},\boldsymbol{\sigma})$, with $\boldsymbol{\mu}=[0.8,3]^T$ and $\boldsymbol{\sigma}=[0.14,3]^T$. Unlike in test case 1 the geometry is fixed here, therefore there is no need to resample the flowfields onto a base grid. For the scalar field variable $f_{i,m}$ we consider the local Mach number $Ma$. 

\subsection{Polynomial ridge function implementation}
\label{sub:ridge_imp}

To obtain dimension reducing ridges, we use the \textit{equadratures} \cite{Seshadri2017} python library which contains an implementation of the polynomial variable projection method described in Section~\ref{sub:poly_ridge}. Since each ridge function is independent of one another, we parallelise the task in an embarrassingly parallel fashion, with computation of individual ridge approximations performed by individual Python worker processes. The code is parallelised using the \textit{joblib} library with the loky backend. Each parallel Python process is run concurrently on a separate computational core on a F72s\_v2 virtual machine (36 physical CPU's, 144 GiB memory) on the Microsoft Azure cloud computing service. For the $N=28.6\times10^3$ points in test case 1, with 500 training designs, training takes approximately 20 minutes per field variable ($\approx 24$ ridge approximations per second). Note that training times can be substantially reduced by taking advantage of spatial correlation in the manner described in Section~\ref{sub:spatial_covar}.

The flowfield reconstruction procedure described in Section~\ref{sub:spatial_covar} is entirely made up of linear algebra operations, thus it is straightforward to implement efficiently with existing linear algebra libraries. For this we use the numpy \cite{Harris2020} python library with Intel\textsuperscript{\tiny\textregistered} MKL. To limit computation and storage requirements, we compute only the $\mK_{12}$ and $\mK_{22}$ blocks of the covariance matrix in \eqref{eqn:covar_blocks}, since only these blocks are required by the Schur complement step in \eqref{eqn:schur}. If further cost savings are required, low-rank approximations for $\mK$ can be obtained, as is done in\cite{scillitoe2021instantaneous}.

\subsection{Convolutional neural network implementation} 
\label{sub:CNN_imp}

\begin{figure}[ht]
\centering
\includegraphics[width=\linewidth]{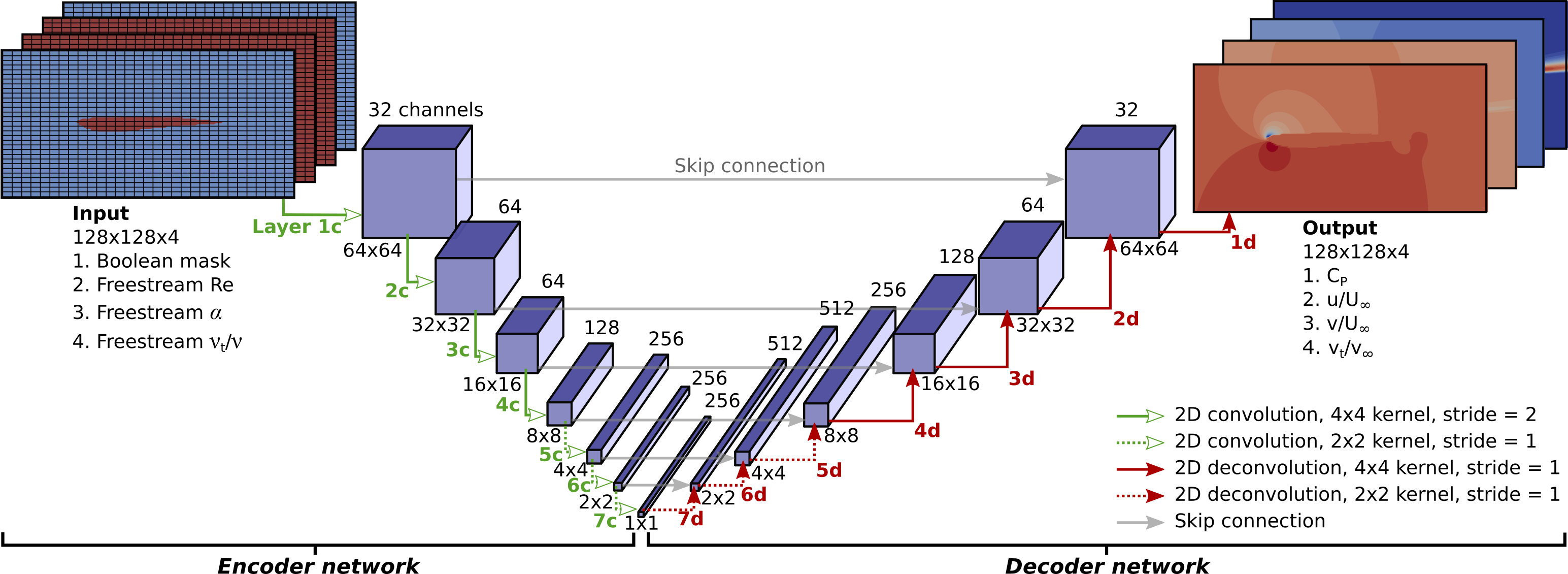}
\caption{The 488k parameter U-net convolutional neural network architecture used for flowfield predictions. Arrows indicate the direction of forward operations. The network is fully convolutional with 14 layers. Further details of the architecture are given in Appendix~\ref{appdx:CNN}.}
\label{fig:CNN_schematic}
\end{figure}

\citet{Thuerey2020} explored a number of CNN architectures for flowfield predictions, and found the U-Net architecture \cite{Ronneberger2015} to be the most successful. In this paper we implement a modified version of the framework proposed by \citet{Thuerey2020}, shown in Figure~\ref{fig:CNN_schematic}. The U-Net architecture consists of an encoder, which progressively down-samples the $128\times128\times4$ input data with strided convolutions. The four input channels consist of a boolean mask to define the aerofoil geometry, and three uniform input channels defining the freestream conditions $Re$, $\alpha_\infty$, and $(\nu_t/\nu)_\infty$. This allows the network to extract increasingly large-scale and abstract information as the number of feature channels grows, until we are left with a $1\times1\times256$ derived feature vector. The decoder then does the opposite, with depooling layers reducing the number of features while increasing the spatial resolution. Eventually we are left with four $128\times128$ output channels, consisting of flowfield estimates for $C_p$, $u/U_\infty$, $v/U_\infty$ and $\nu_t/\nu_\infty$. Skip connections help the network to consider low-level input information during the reconstruction of the solution in the decoding layers. The weights $w_{ij}^{(l)}$ (and biases $b_i^{(l)}$d) of the neural network are obtained by minimising the loss function
\begin{align} \label{eqn:CNN_loss}
\begin{split}
\mathcal{L} =&  \frac{1}{M}\frac{1}{N} \sum_{m=1}^M \sum_{i=1}^N \phi_{hub}\left(\hat{C_p}_{i,m} - {C_p}_{i,m} \right) + \phi_{hub}\left(\frac{\hat{u}_{i,m}}{U_\infty} - \frac{u_{i,m}}{U_\infty}\right) \\
& + \phi_{hub}\left(\frac{\hat{v}_{i,m}}{U_\infty} - \frac{v_{i,m}}{U_\infty}\right)  + \phi_{hub}\left(\frac{\hat{\nu_t}_{i,m}}{\nu_\infty} - \frac{{\nu_t}_{i,m}}{\nu_\infty}\right),
\end{split}
\end{align} 
with $\hat{\cdot}$ denoting the CNN prediction for the field variable, and $\phi_{hub}$ denoting the Huber loss function\cite{Huber1964}. As seen in \eqref{eqn:CNN_loss}, the error is averaged over all $M$ number of training designs and $N$ number of grid points. This loss function is differentiable, enabling it to be back-propagated into the network in order to compute the weight gradient $\nabla_{\mathbf{w}}\mathcal{L}$, which is important for the optimisation (\textit{learning}) process.

The complexity of the above CNN is altered in this paper by varying the number of channels (and therefore the number of  parameters), with the 488k parameter CNN in Figure~\ref{fig:CNN_schematic} found to offer the best compromise between training times and accuracy. All CNN's detailed in this paper are implemented in the PyTorch python library, and trained on a GTX970 1664 core NVIDIA GPU. Training times range from 50 seconds to 80 minutes, depending on the complexity of the network and the number of training designs. Further details are included in Appendix~\ref{appdx:CNN}.

\section{Test case 1: Design of a Subsonic Aerofoil}

\begin{figure}[t] 
\centering
\includegraphics[width=\linewidth]{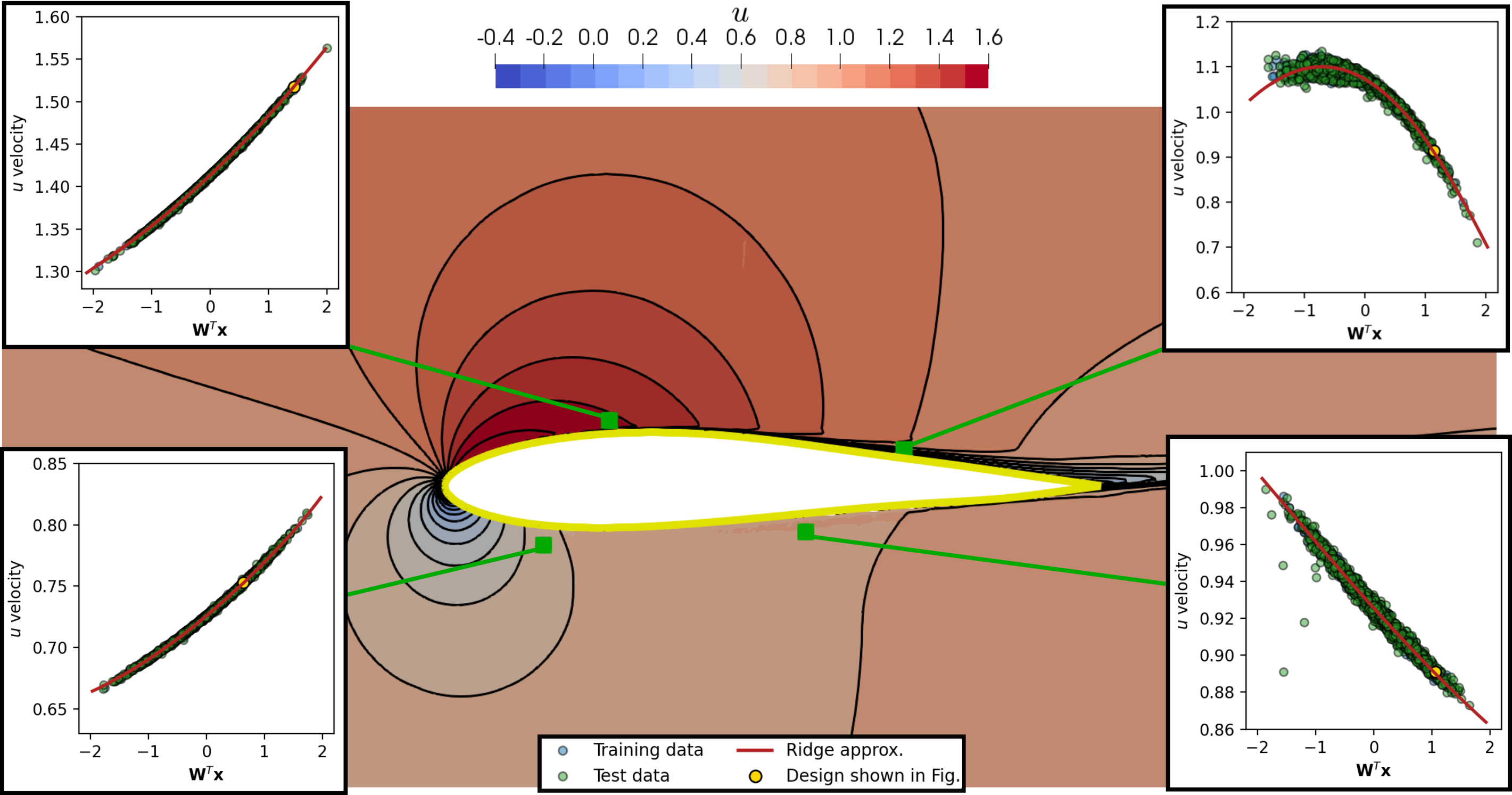}
\caption{Normalised axial velocity, $u/U_\infty$, for a deformed aerofoil (from the test set) at an angle of incidence of $\alpha_\infty=10^{\circ}$. Colour contours show predictions from the second degree polynomial ridge approximations, isolines show the true CFD solution. Sufficient summary plots at four locations are shown.}\label{fig:AoA10_u}
\end{figure}

As an initial demonstration, ridge approximations are obtained for the static pressure coefficient $C_p=({p_0}_\infty-p_0)/({p_0}_\infty-p_\infty)$, normalised axial velocity $u/U_\infty$, and turbulent viscosity ratio $\nu_t/\nu$, with 500 aerofoil designs used for training. Flowfield predictions for a new aerofoil design can then be obtained by transforming the new design vector ${\mathbf{x}}$ to the reduced dimensional space, and evaluating the polynomial ridge approximation $p_{i} \left( \mW_{i}^{T} \vx ; \vs_{i} \right)$ at each point $i=1,\dots,N$ in the flowfield. In Figure~\ref{fig:AoA10_u}, the predicted normalised $u$ velocity is shown for a randomly selected design from the test set. Qualitatively, the predictions are in close agreement to the \emph{true} velocity field from the CFD solution, shown by the black iso-lines. The design shown here was not used during training, hence the ridge approximations appear to be able to \emph{learn} enough of the flow physics to be able to make predictions for the new design. 

In addition to making flowfield predictions, a second key aspect of the ridge approximations are \emph{sufficient summary plots}, seen in the insets within Figure~\ref{fig:AoA10_u}. Applied to aerofoil design, they provide visualisation of how the quantity of interest, in this case the $u$ velocity at selected locations, is altered as the shape is changed. This is made possible since the sufficient summary plot shows the $u$ velocity over the reduced dimensional space $\mathbf{W}^T\mathbf{x}\in\mathbb{R}$, instead of the full space $\mathbf{x}\in\mathbb{R}^{50}$. The low scatter of train and test points in the summary plots highlights that low dimensional structure has been successfully found.

\subsection{Accuracy of Ridge Approximations}
\label{sub:accuracy}

\begin{figure}[ht]
	\centering
     \includegraphics[width=0.3\linewidth]{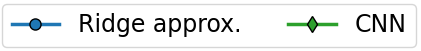} \vspace{-8pt} \\
     \subfloat[$C_p$, $\alpha_\infty=0^{\circ}$
  		\label{subfig:MAE_0_Cp}]{
  		\includegraphics[width=0.35\linewidth]{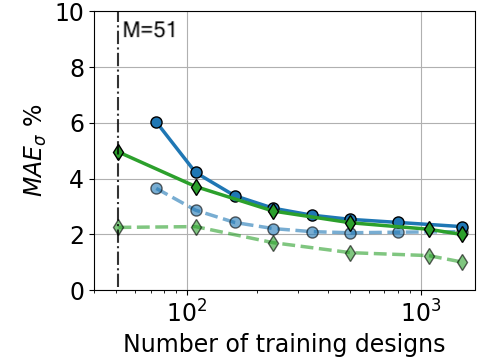}}   \hfil
  	\subfloat[$u/U_\infty$, $\alpha_\infty=0^{\circ}$
  		\label{subfig:MAE_0_u}]{%
  		\includegraphics[width=0.35\linewidth]{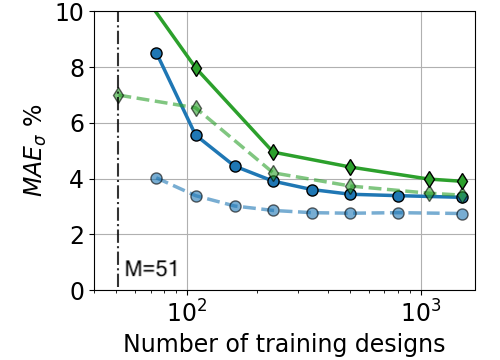}}   \\
     \subfloat[$C_p$, $\alpha_\infty=10^{\circ}$
  		\label{subfig:MAE_10_Cp}]{%
  		\includegraphics[width=0.35\linewidth]{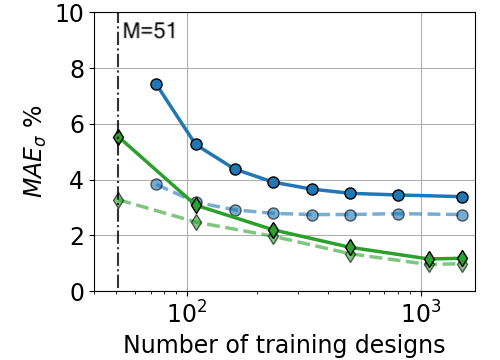}}   \hfil
  	\subfloat[$u/U_\infty$, $\alpha_\infty=10^{\circ}$
  		\label{subfig:MAE_10_u}]{%
  		\includegraphics[width=0.35\linewidth]{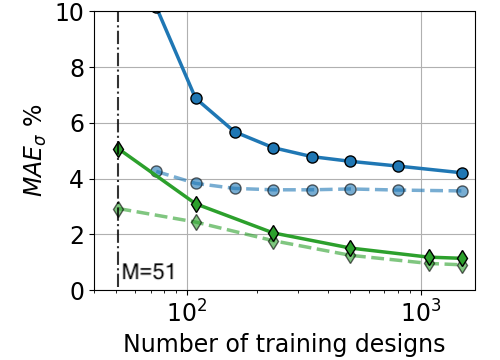}}
	\caption{Normalised MAE of $C_p$ and $u/U_\infty$ predictions for first degree polynomial ridge approximations and convolutional neural network at two angles of incidence. Solid lines = test error, dashed lines = training error. Results are averaged over three randomly selected train/test splits.}
	\label{fig:MAE_vs_N}
\end{figure}

To explore how the accuracy of the proposed ridge approximation framework compares to a state-of-the-art convolutional neural network (CNN), ridge approximations and the 488k parameter U-Net in Figure~\ref{fig:CNN_schematic} are trained on $C_p$ and $u/U_\infty$ around the aerofoil at two angles of incidence, $\alpha_\infty=0^{\circ}$ and $\alpha_\infty=10^{\circ}$. To enable fair accuracy comparisons the ridge approximations are obtained on the same $128^2$ grid used for the CNN instead of the finer 90x318 grid used elsewhere. The normalised mean absolute error of predictions 
\begin{equation} \label{eqn:MAE}
MAE_{\sigma} =  \frac{1}{N}\sum_{i=1}^N \left[ \frac{\frac{1}{M}\sum_{m=1}^M \abs{\hat{f}_i(\mathbf{x}_m)-f_i(\mathbf{x}_m)}}{\sigma[f_i(\mathbf{x}_m)]}  \right],
\end{equation}
is then measured for the training and test designs. At each point, the absolute error is averaged for $M$ number of designs, then normalised by the standard deviation of the true data at that point. The result is averaged over all $N$ number of points. This is done whilst varying the number of training designs, in order to obtain the error curves in Figure~\ref{fig:MAE_vs_N}. Both the ridge approximation and CNN prediction errors exhibit the same general trends; for low numbers of training designs the test errors are considerably higher than the training errors, suggesting a lack of generalisation (i.e. the models are overfitting to the training data). However the test errors are significantly decreased as the number of training designs is increased. Perhaps surprisingly, despite the data hungry reputation of deep neural networks, the CNN framework is able to achieve low accuracies with a relatively small number of training designs ($MAE_{\sigma}<10\%$ with $M_{train}<100$). As noted by \citet{Ronneberger2015}, this is a key benefit of the U-Net and other CNN architectures; the strong use of data augmentation via convolutions allows the available training data to be used more efficiently (compared to a fully connected deep neural network). 

\begin{figure}[ht]
	\centering
     \includegraphics[width=0.35\linewidth]{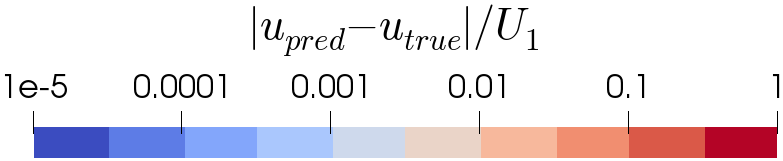} \\
     \subfloat[Ridge approximations
  		\label{subfig:uerror_ridge}]{%
  		\includegraphics[width=0.48\linewidth]{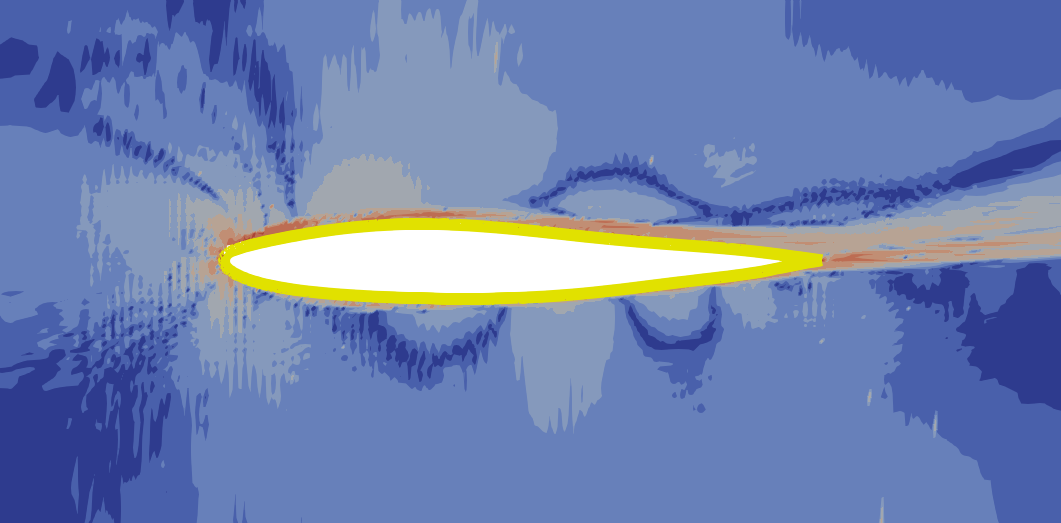}}   \hfil 
     \subfloat[Convolutional neural network
  		\label{subfig:uerror_CNN}]{%
  		\includegraphics[width=0.48\linewidth]{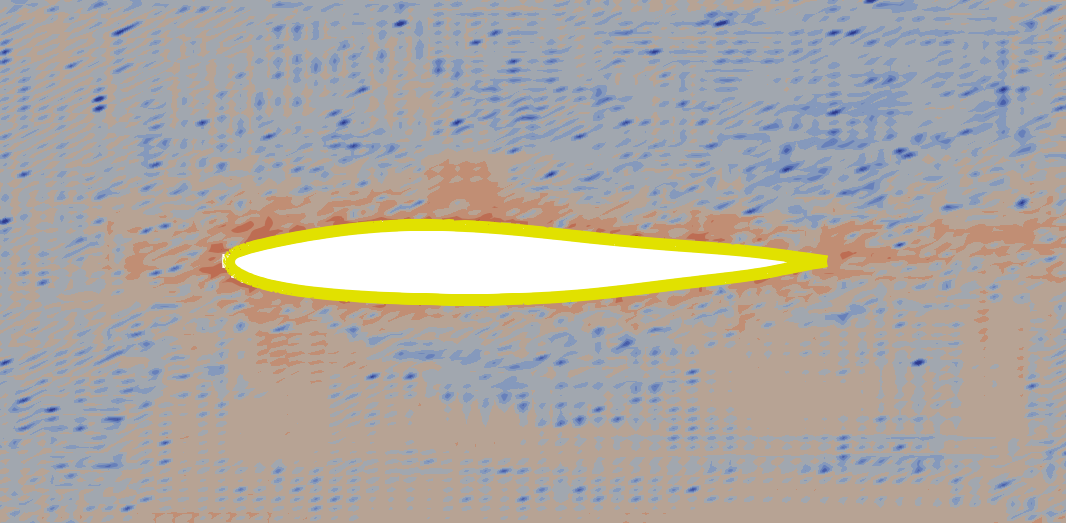}}
	\caption{Comparison of absolute error in predictions of normalised axial velocity from the first degree ridge approximations and convolutional neural network, for an aerofoil from the test set at $\alpha_\infty=10^{\circ}$.}
	\label{fig:uerror}
\end{figure}

Encouragingly, for the $\alpha_\infty=0^{\circ}$ case (Figs~\ref{subfig:MAE_0_Cp} and~\ref{subfig:MAE_0_u}) the ridge approximations are able to achieve mean test errors which are competitive with the CNN. At the higher angle of incidence of $\alpha_\infty=10^{\circ}$ (Figs~\ref{subfig:MAE_10_Cp} and~\ref{subfig:MAE_10_u}) the flowfield is more complex, with stronger streamline curvature leading to larger pressure gradients and greater non-linear behaviour. This case is more challenging for the ridge approximations, with test errors of $MAE_{\sigma}\approx4\%$ compared to the CNN test errors of $MAE_{\sigma}\approx1\%$. However, accuracies of under $4\%$ would be acceptable for many preliminary design applications. Additionally, as the MAE fields in Figure~\ref{fig:uerror} show, the ridge approximations' prediction errors are actually lower than the CNN errors in many regions of the flow. Compared to the rather randomly distributed CNN errors, the ridge approximations' errors exhibit a smoother spatial distribution, with higher errors near to the aerofoil surface and in the wake.  

In the above results, ridge approximations are computed using first degree polynomials. In Table~\ref{tab:mae}, results with higher degree polynomials are presented. Clearly, increasing the polynomial degree, $p$, allows for lower test errors to be obtained (although excessively high-degree polynomials will suffer from over-fitting). However, increasing $p$ also increases the cardinality of the polynomial; if all interaction terms are included, and the polynomial has degree $p$ in all $n$ directions, its cardinality is given by $L = (p+1)^n$. More degrees of freedom result in higher variances of the resulting regression estimators, necessitating more training samples to mitigate this. The exponential dependence on polynomial degree implies that higher degree polynomials are significantly more costly when in higher dimensions. This highlights a key benefit of obtaining polynomials in a dimension reducing subspace of dimension $n$, where $n\ll d$. Various approaches exist to mitigate this exponential scaling\cite{seshadri2017effectively}. However, even then, it is still desirable to reduce the number of dimensions. As a compromise between accuracy and computational cost, second degree polynomials are used in the remainder of the results presented for the present test case. 

\begin{table}[ht] 
\centering
\caption{Normalised MAE of ridge approximations with various polynomial degrees, for the $C_p$ and $u/U_{\infty}$ fields at an angle of incidence of $\alpha_\infty=10^{\circ}$.}\label{tab:mae}
\begin{tabular}{ccccc} 
\toprule
\multirow{2}{*}{Polynomial degree, $p$} & \multicolumn{2}{c}{Training $MAE_{\sigma} \; (\%)$}& \multicolumn{2}{c}{Test $MAE_{\sigma} \; (\%)$}  \\
\cline{2-5}
& \hspace{12pt}$C_p$\hspace{12pt} & $u/U_{\infty}$ & \hspace{12pt}$C_p$\hspace{12pt} & $u/U_{\infty}$   \\
\midrule
1 & $2.8$ & $3.5$ & $3.6$ &  $4.5$  \\
2 & $2.1$ & $2.8$    &  $2.9$ & $3.7$   \\
3 & $1.6$  & $2.0$  &   $2.7$ & $3.6$ \\
\bottomrule
\end{tabular}
\end{table}


\subsection{Design Space Exploration} 

\begin{figure*}[ht]
\centering
\includegraphics[width=\linewidth]{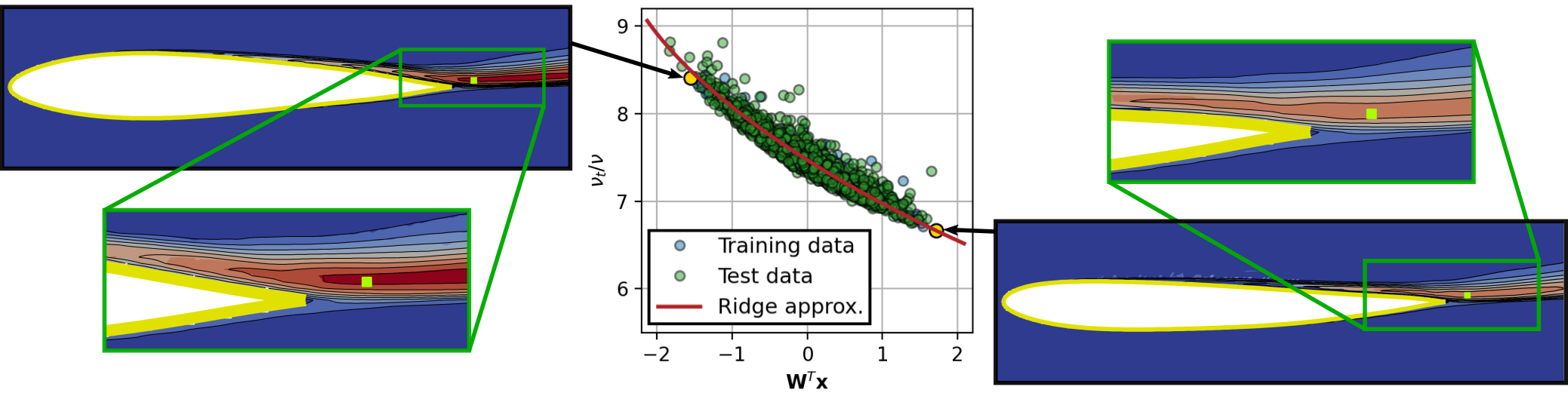}
\caption{Turbulent viscosity ratio, $\nu_t/\nu_\infty$, for two deformed aerofoils (from the test set) at an angle of incidence of $\alpha_\infty=10^{\circ}$. Colour contours show predictions from the ridge approximations, isolines show the true CFD solutions. The two designs are at either end of the sufficient summary plot for a point in the wake.}
\label{fig:AoA10_nut}
\end{figure*}

The dimension reducing nature of the ridge approximations is particularly useful for design space exploration. To demonstrate, ridge approximations for the turbulent viscosity ratio $\nu_t/\nu_\infty$ are obtained using all 500 training designs. In Figure~\ref{fig:AoA10_nut} a sufficient summary plot for a point in the aerofoil's wake is shown. The turbulent viscosity ratio in the wake is of interest because high levels of turbulence in wakes can be a leading source of loss in many flows \cite{Scillitoe2016a}. The sufficient summary plot is examined, and a design at either end of the $\mathbf{W}^T\mathbf{x}$ subspace is selected. Contours of the turbulent viscosity ratio for the two designs are shown in Figure~\ref{fig:AoA10_nut}. Through this approach, the end user can explore how flowfield variables at select locations in the flowfield change as they traverse the design space. Such an activity would be far more challenging when negotiating the original design space $\mathbf{x}\in\mathbb{R}^d$. Effectively, dimension reduction distils the design space into critical directions along which quantities of interest exhibit maximum variance, which enables efficient design space exploration.

\begin{figure}[ht]
	\centering
     \subfloat[
  		\label{subfig:inactive_0}]{%
  		\includegraphics[width=0.49\linewidth]{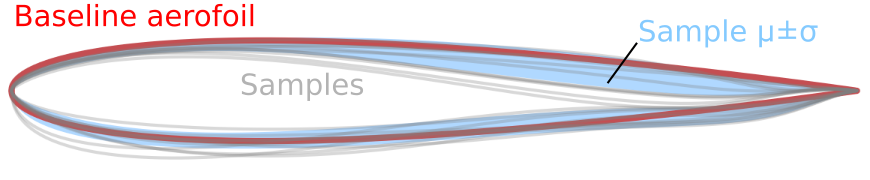}}   \hfil 
     \subfloat[
  		\label{subfig:inactive_2}]{%
  		\includegraphics[width=0.49\linewidth]{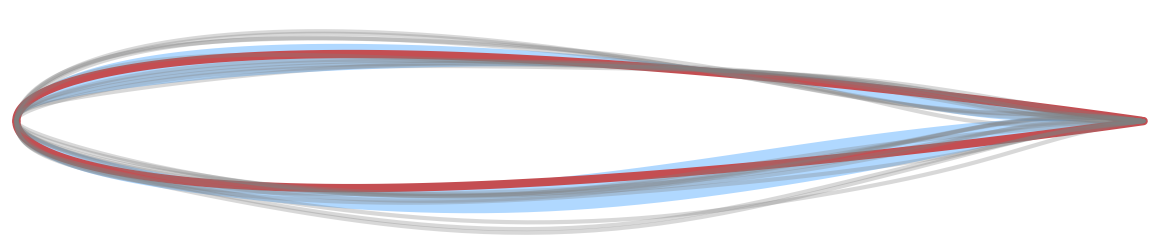}}
	\caption{Design envelopes obtained by sampling the inactive subspaces of the ridge approximations for $u/u_\infty$ at locations a) and b) in Figure~\ref{fig:AoA10_u}. 1000 samples are generated using a hit-and-run algorithm \cite{Constantine2016}, with the grey lines corresponding to the blade profiles of 10 randomly selected samples.}
	\label{fig:inactive}
\end{figure}

In addition to exploring directions of maximum variance, one can also explore directions along which quantities of interest exhibit minimum variance. Following Section~\ref{sub:ridges}, the input design vector can be decomposed as
\begin{equation}
\begin{split}
\vx &= \mW \mW^T\vx + \mV \mV^T\vx \\
& = \mW\vw + \mV\vv
\end{split}
\end{equation}
where $\vw = \mW^T\vx$ is the active coordinate and $\vv = \mV^T\vx$ the inactive coordinate. The inactive subspace $\mV$ permits identification of designs which are invariant with regards to the quantity of interest. The task is to find input vectors $\vx$ with a fixed $\vw$ but different $\vv$, whilst obeying the constraint $-1\le\vx\le1$, for which we use a hit-and-run algorithm similar to that implemented in \cite{Constantine2016}. In Figure~\ref{fig:inactive}, this is done for the ridge approximations at the locations labelled a) and b) in Figure~\ref{fig:AoA10_u}. The resulting \textit{design envelopes} inform us how the aerofoil can be deformed without the quantity of interest $u/u_{\infty}$ at these two locations being affected. Such information can be used to form manufacturing tolerances, or to decide whether to scrap components, as is done in \cite{scillitoe2020design,wong2021blade,wong2020blade}.

\subsection{Physical Insight}
\label{sub:physics}

In addition to exploring the design space, ridge functions can also provide important physical insights. As an example, we consider the stagnation pressure around the aerofoil. In an ideal (inviscid) incompressible flow, the stagnation pressure, $p_0 = p + 0.5 \rho U^2$, is constant along streamlines\footnote{The stagnation pressure is constant everywhere if the flow is inviscid and irrotational.}. Streamline curvature results in a reversible exchange between the static pressure $p$ and dynamic pressure $0.5\rho U^2$. However, in a real flow, irreversible processes such as viscous effects cause a loss in $p_0$, which it is important to control. Take the case of an aircraft; high losses over the wing are synonymous with high drag \cite{Wang2018}, while in the engines, high losses lead to decreased component efficiencies, both of which lead to a higher fuel consumption. The pressure loss at a given location $\vs$ can be quantified through the loss coefficient
\begin{equation} \label{eqn:Yp}
Y_p(\vs) = \frac{{p_0}_\infty-p_0(\vs)}{0.5 \rho_\infty U_\infty^2},
\end{equation}
where ${p_0}_\infty$ is the freestream stagnation pressure, and $Y_p=0$ indicates no loss in stagnation pressure, while $Y_p=1$ indicates a total loss in stagnation pressure. 

\begin{figure*}[ht]
\centering
\includegraphics[width=\linewidth]{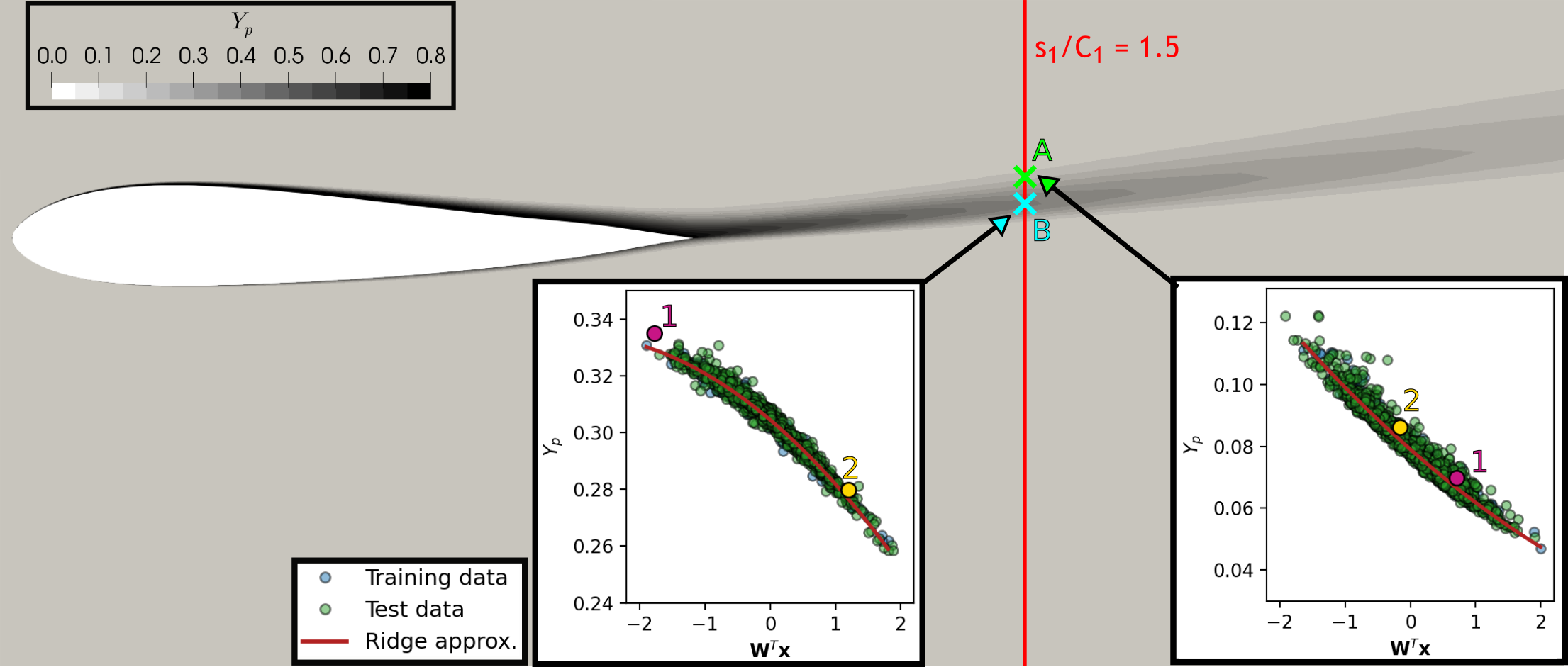}
\caption{Contours of loss coefficient, $Y_p$, for a deformed aerofoil from the test set (``Design 1'') at an angle of incidence of $\alpha_\infty=10^{\circ}$. A line at 50\% axial chord downstream of the trailing edge is highlighted, and sufficient summary plots are shown for two points within the wake, on this line. Design 1, and a second design, are highlighted in the summary plots.}
\label{fig:Yp_contours}
\end{figure*}

\begin{figure}[ht]
	\centering
	\subfloat[
  		\label{subfig:Yp_profile}]{%
  		\includegraphics[height=0.30\linewidth]{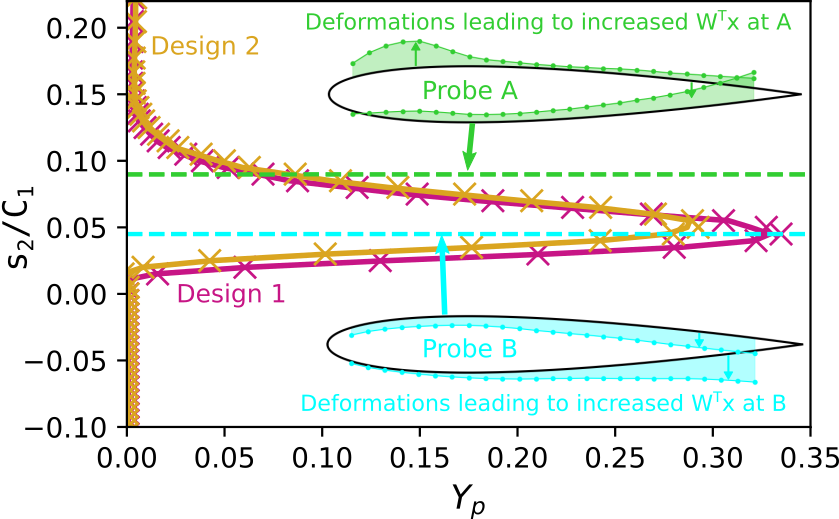}}   \hfil 
	\subfloat[
  		\label{subfig:Yp_ridge}]{%
  		\includegraphics[height=0.30\linewidth]{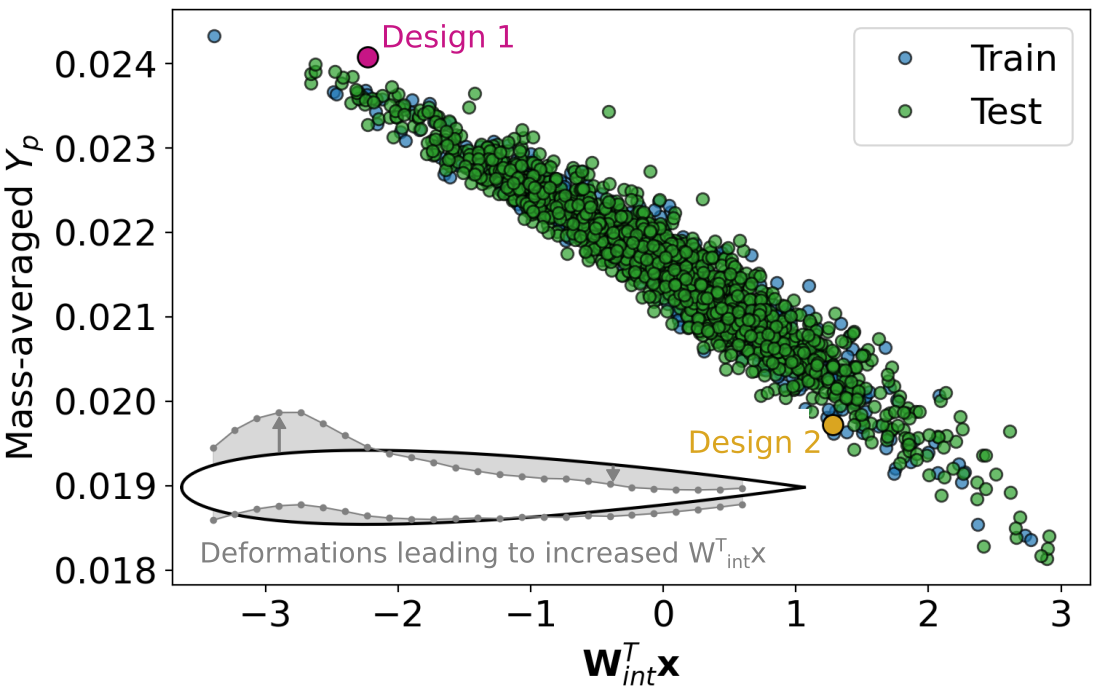}}
  	\caption{a) Profiles of loss coefficient $Y_p$ across $s_1/C_1=1.5$, and their ridge approximations, for the two designs highlighted in Figure~\ref{fig:Yp_contours}. Also shown are the $s_2$ locations of probes A and B from Figure~\ref{fig:Yp_contours} (dashed lines), along with the projections of $\mathbf{W}$ at these two locations onto the baseline aerofoil. b) Sufficient summary plot for the mass-averaged loss coefficient across the $s_1/C_1=1.5$ line, with designs 1 and 2 highlighted, and the associated subspace vector $\mathbf{W}_{int}$ projected onto the baseline aerofoil.}
     \label{fig:Yp_local_int}
\end{figure}

The loss coefficient $Y_p$ is calculated from the CFD data for 500 training designs, and the approximations for a randomly selected test design are visualised in Figure~\ref{fig:Yp_contours}. In Figure~\ref{subfig:Yp_profile}, profiles of $Y_p$ across the $s_1/C_1=1.5$ plane are plotted for the two designs labelled 1 and 2 in Figure~\ref{fig:Yp_contours}. For both designs, the high loss in the wake (around $0\le s_2/C_1 \le 0.15$) is accurately predicted by the ridge functions. 

The sufficient summary plots in Figure~\ref{fig:Yp_contours} show how $Y_p$ at probe locations A and B change as the design space is traversed. Furthermore, the dimension reducing subspace vector $\mathbf{W}$ can be examined. Recalling that in the present framework $\mathbf{x}$ contains the Hicks-Henne bump amplitudes, the elements of $\mathbf{W}$ show us how the bump amplitudes affect the reduced coordinates $\mathbf{W}^T\mathbf{x}$. In Figure~\ref{subfig:Yp_profile}, the scaled elements of $\mathbf{W}$ for the $Y_p$ ridge functions at A and B are projected onto the baseline aerofoil geometry\footnote{Each element of the subspace weight vector is plotted at the surface location of its associated bump.}. The weights offer important physical insights here; they show us how the aerofoil surfaces must be deformed in order to affect the loss coefficient $Y_p$ at probes A and B.

\subsection{Obtaining Global Quantities}
Understanding the sensitivity of local flow variables to a design can be invaluable to a designer. But, sometimes, a designer may also be interested in global quantities of interest. For example, in the present case, the mass averaged loss across the $s_1/C_1=1.5$ line in Figure~\ref{fig:Yp_contours} is an important quantity, and is given by
\begin{equation} \label{eqn:Yp_int}
\bar{Y}_p = \frac{\int u(s_1,s_2) Y_p(s_1,s_2) ds_2}{\int u(s_1,s_2) ds_2}, 
\end{equation}
where $u(s_1,s_2)$ and $Y_p(s_1,s_2)$ denote the local axial velocity and loss coefficient respectively. The mass averaged loss represents the total pressure loss caused by the aerofoil. A surrogate model for $\bar{Y}_p$ could be obtained by preprocessing the CFD data, and then computing a ridge approximation for the precomputed $\bar{Y}_p$ values. However, alternatively, the local ridge functions themselves can be integrated following the approach outlined in Appendix~\ref{appdx:integrate_ridge}. A dimension reducing subspace for $\bar{Y}_p$, $\mathbf{W}_{int},$ is obtained directly from the ridge approximations for $u(s_1,s_2)$ and $Y_p(s_1,s_2)$. The resulting one dimensional sufficient summary plot for $\bar{Y}_p$ is shown in Figure~\ref{subfig:Yp_ridge}. It is apparent that designs 1 and 2 lie at either end of the design space with regards to their mass averaged loss, indicating that design 2 is objectively the better design (with regards to loss). Similarly to before, in Figure~\ref{subfig:Yp_ridge} the elements of $\mathbf{W}_{int}$ are projected onto the baseline aerofoil. This shows that the total loss can be reduced by deforming the suction surface outwards near the leading edge, and inwards near to the trailing edge. 

\begin{figure}[ht]
	\centering
     \includegraphics[width=\linewidth]{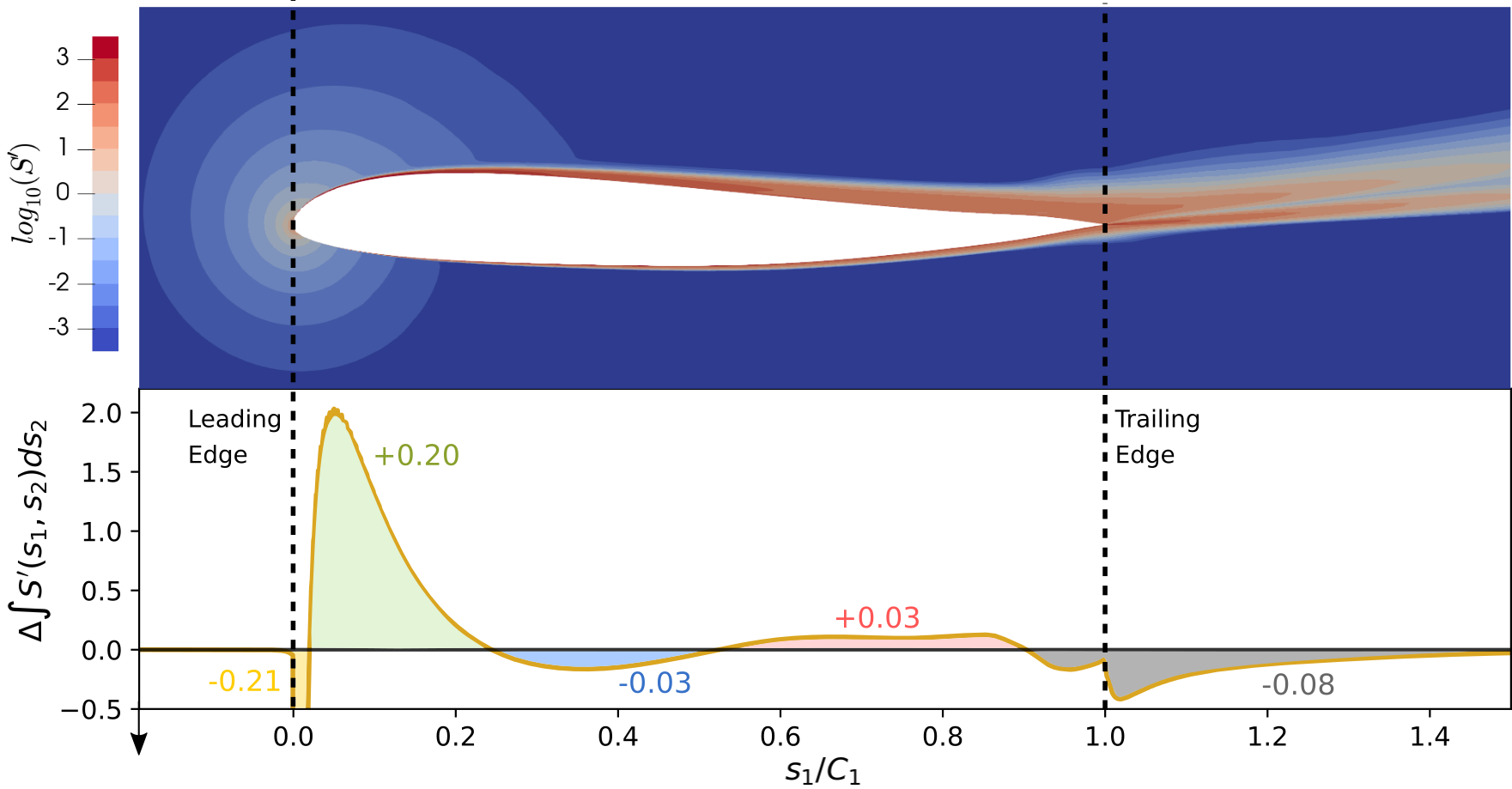} 
	\caption{Contours of total entropy generation rate, $S^{\prime}$, for design 2 (see Figs.~\ref{fig:Yp_contours}-~\ref{fig:Yp_local_int}), and axial profile of the difference in y-averaged generation rate between design 2 and the baseline design. The areas under the profile at various locations are also shown. The units of $S^{\prime}$ are $Wm^{-3}K^{-1}$.}
	\label{fig:Sgen}
\end{figure}

To understand why design 2 has a lower loss, a second law analysis can be performed. Herwig and Schmandt \cite{Herwig2014} show that the overall entropy generation rate
\begin{equation} \label{eqn:Sgen}
S^{\prime} = \frac{\rho(\nu+\nu_t)}{T} \left(\frac{\partial u_i}{\partial s_j} + \frac{\partial u_j}{\partial s_i} \right) \frac{\partial u_i}{\partial s_j},
\end{equation}
can be used to understand the sources of drag or pressure loss within a flow. The contours of $S^{\prime}$ in Figure~\ref{fig:Sgen} show that, at $\alpha=10^{\circ}$, the majority of the entropy generation comes from the suction surface boundary layer and the wake region\footnote{Decomposing Equation~\ref{eqn:Sgen} into entropy generation due to laminar viscosity $\nu$, and that due to turbulent viscosity $\nu_t$, shows that laminar irreversibilities dominate over the aerofoil surfaces, while turbulent mixing dominates in the wake.}. In Figure~\ref{fig:Sgen} the delta between the $y-$averaged entropy generation for design 2 and the baseline design ($\mathbf{W}^T\mathbf{x}=0$) is shown. Deforming the suction surface outwards near the leading edge, and flattening the aft portion (see Fig.~\ref{subfig:Yp_ridge}), appears to increase entropy generation over the first quarter of the suction surface. But, this is offset by a reduction in entropy generation in the wake and leading edge stagnation region, leading to a net decrease in the entropy generation. Whilst an experienced aerodynamicist may have foreseen this conclusion, the present example serves to illustrate the potential of an easily interpretable surrogate model which can be readily integrated to obtain global quantities. 

\subsection{Exploiting Spatial Correlation}
\label{sub:spatial_covar}

\begin{figure}[ht]
	\centering
	\includegraphics[width=0.8\linewidth]{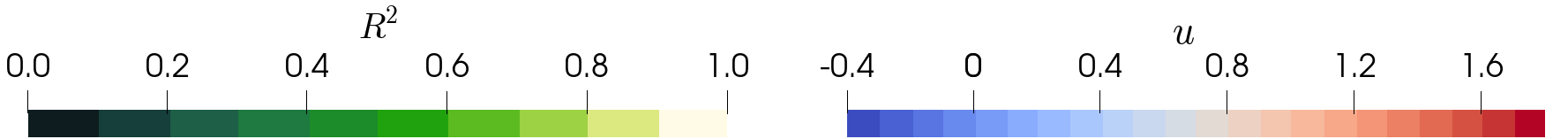} \vspace{2pt}\\
  	\includegraphics[width=0.99\linewidth]{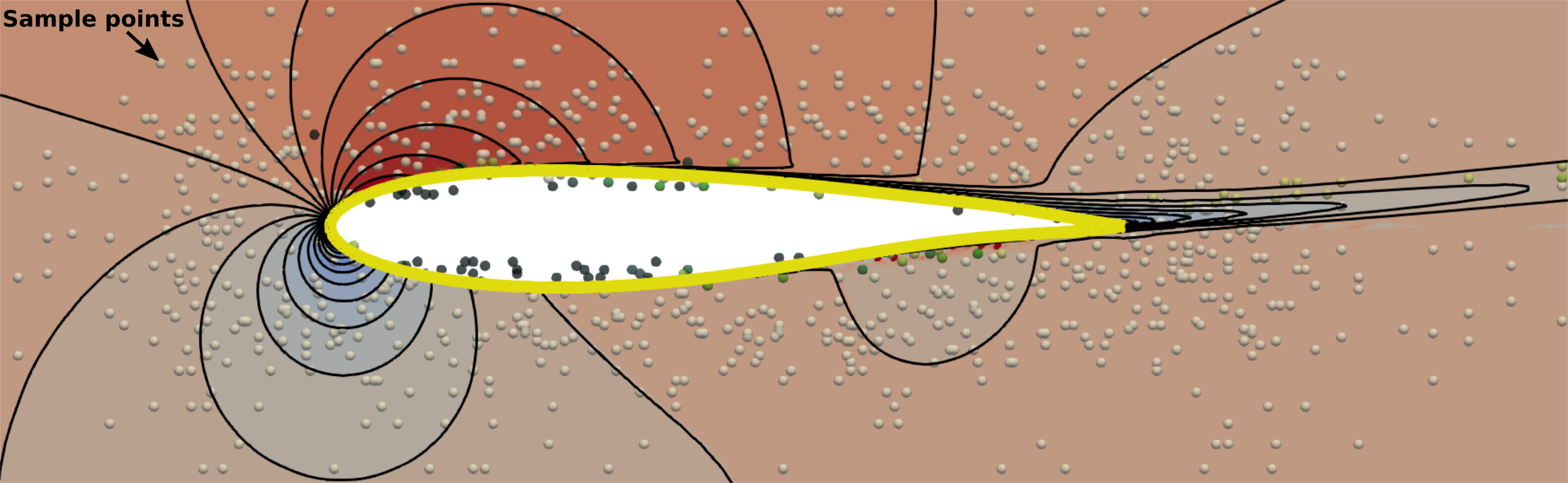}
  	\caption{Flowfield estimate of axial velocity, $u/U_\infty$, obtained by upsampling ridge approximations at $J=1000$ randomly subsampled points. The aerofoil is from the test set, and the angle of incidence is $\alpha_\infty=10^{\circ}$. Colour contours show upsampled ridge approximations, while the iso-lines show the true CFD solution. The subsampled points are coloured by the mean $R^2$ score of their ridge approximations over the training data.} 
     \label{fig:aerofoil_upsample}
\end{figure}

Up to this point, flowfield predictions have been made using ridge approximations at every grid point. For larger grids,  such as three dimensional ones, obtaining $N$ number of ridge approximations might be computationally intractable. Taking a similar approach to how the CNN takes advantage of spatial coherence in the flowfield, the method outlined in Section~\ref{sub:covar} explots spatial correlations to reduce the number of ridge approximations required. In Figure~\ref{fig:aerofoil_upsample}, ridge approximations are obtained for $J=1000$ randomly subsampled points (a sampling rate of $J/N=3.5\%$). Since the baseline mesh has a high mesh density near the aerofoil (where the gradients are high), the subsampled points are also clustered near to the aerofoil, which is desirable. The covariance matrix $\mK$, computed on the training data, is then used to \textit{upsample} the ridge approximations at the subsampled points back to the remaining $N-J$ points.

%

The colouring of the subsampled points in Figure~\ref{fig:aerofoil_upsample} shows that there are a small number of points with poor quality ridge approximations. Preliminary work\cite{scillitoe2021instantaneous} showed that these poor quality approximations can cause spurious noise in the flowfield estimates when upsampled. As discussed in Section~\ref{sub:covar}, the training $R^2$ score can be used as a heuristic to measure the quality of each ridge. Ridges with $R^2<0.6$ are ignored during upsampling, with their grid points added back into the $N-J$ set of points. As seen in Figure~\ref{fig:aerofoil_upsample}, this strategy results in qualitatively accurate flowfield predictions using ridge approximations at only 3.5\% of the grid points, reducing the training time by a factor of approximately 29.
\section{Test case 2: Varying Freestream Conditions for a Transonic Wing}
\label{sec:results2}

Following on from the success of the first test case, we now turn to the transonic wing test case described in Section~\ref{sub:onera} in order to expose the limitations of polynomial ridge functions. Here the geometry, represented by $\vx \in \mathbb{R}^{b}$, is fixed. However, the boundary conditions, represented by $\vv \in \mathbb{R}^{b}$, are varied. The transonic nature of the flow, and the possibility of extrapolating beyond the training dataset, make this test case particularly challenging. The input vector $\mathbf{v}$ only consists of two dimensions in this case, $Ma_{\infty}$ and $\alpha_{\infty}$. We could once again obtain one dimensional ridge functions. However, two dimensional ridge functions are still simple to visualise. Considering a ridge function of the form
\begin{equation}
f(\vv) \approx g(\mW^T \vv) \quad \text{where} \quad \mathbf{W}\in \mathbb{R}^{b\times n} \quad \text{and} \quad \mathbf{v} \in \mathbb{R}^b,
\end{equation}
we set $\mathbf{W}=\mathbf{I}$ (the identity matrix), resulting in two dimensional ridge functions ($n=b$). 

\subsection{Interpolation versus extrapolation}

\begin{figure}[ht]
	\centering
	\includegraphics[width=0.6\linewidth]{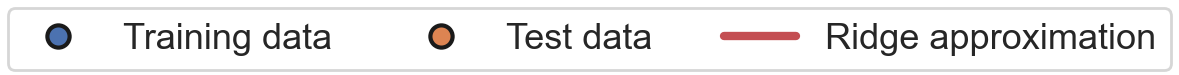} \vspace{-8pt} \\
	\subfloat[DoE point 1: $Ma_{\infty}=1.02,\alpha_{\infty}=5.14^{\circ}$
  		\label{subfig:onera_profle_interp}]{%
  		\includegraphics[width=0.49\linewidth]{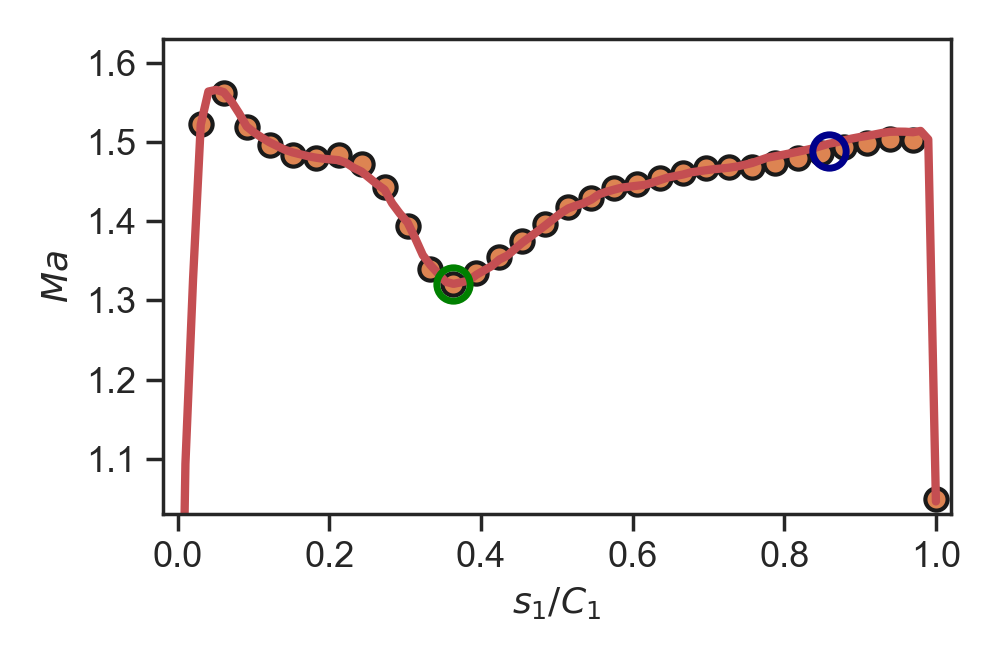}}   \hfil 
	\subfloat[DoE point 2: $Ma_{\infty}=0.72,\alpha_{\infty}=-1.45^{\circ}$
  		\label{subfig:onera_profile_extrap}]{%
  		\includegraphics[width=0.49\linewidth]{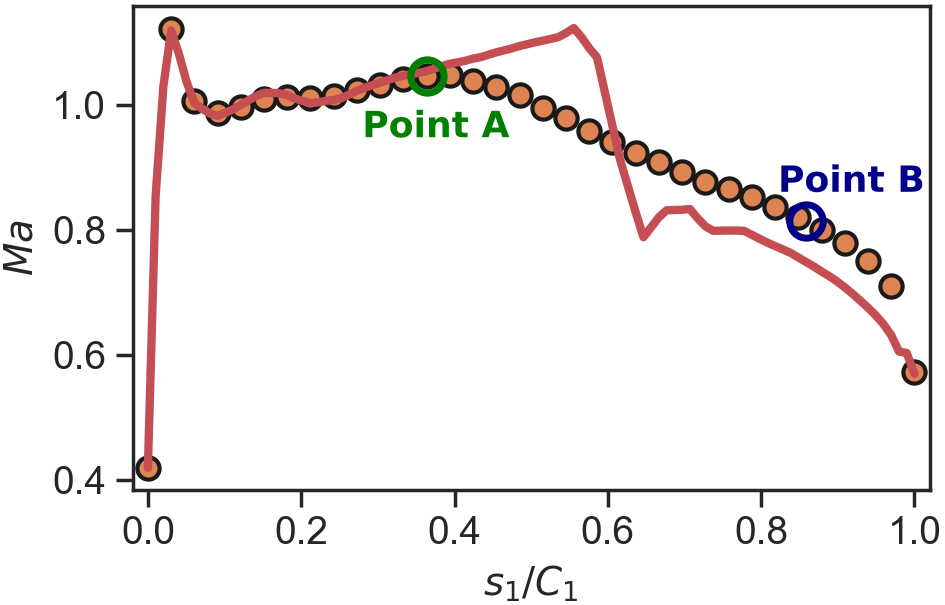}} \\
	\subfloat[Ridge at location A
  		\label{subfig:onera_ridge_A}]{%
  		\includegraphics[width=0.49\linewidth]{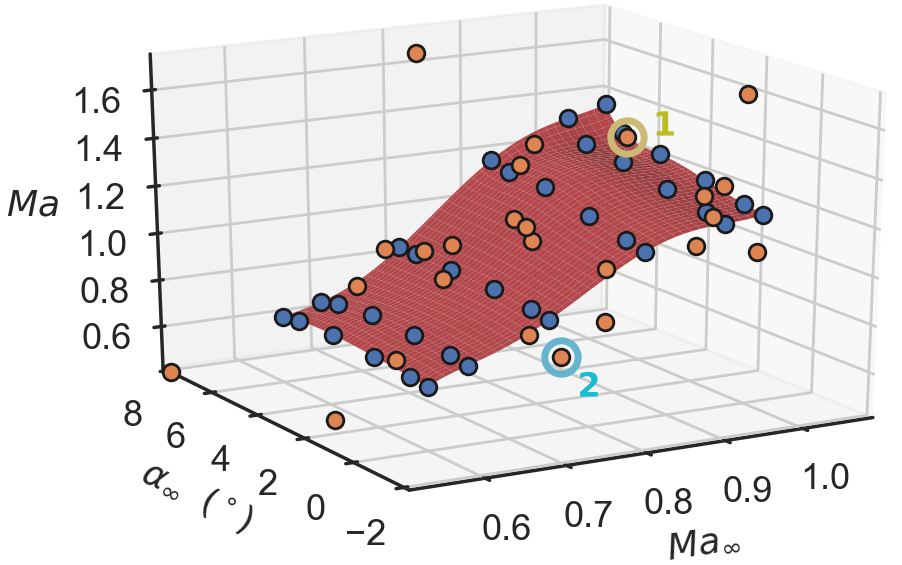}}   \hfil 
	\subfloat[Ridge at location B
  		\label{subfig:onera_ridge_B}]{%
  		\includegraphics[width=0.49\linewidth]{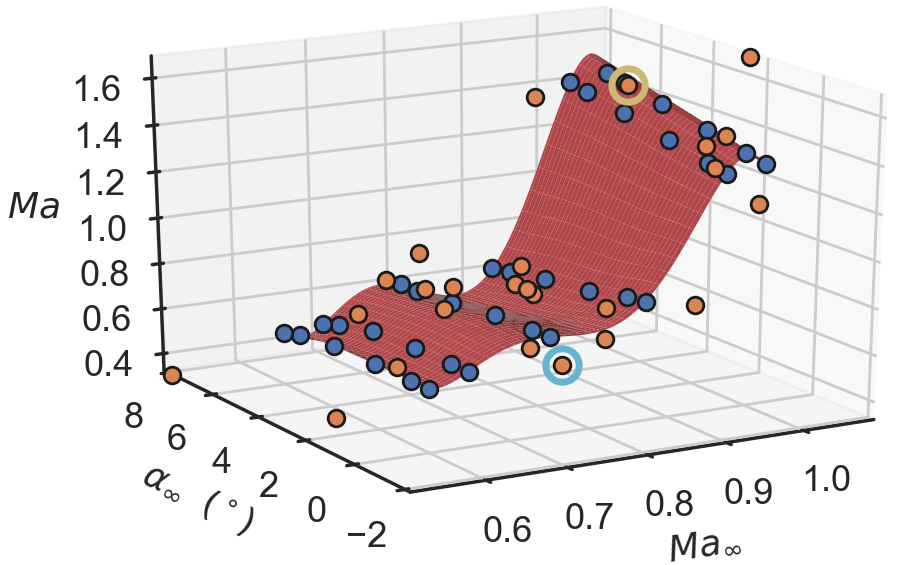}}
  	\caption{Polynomial ridge approximations for Mach number on the surface of the transonic wing at the $s_3/S=0.2$ span-wise slice. The predicted Mach distributions for two DoE points from the \textit{test} set are shown in a) and b), while the corresponding ridges for two $s_1/C_1$ locations are visualised in c) and d). The two DoE points are also labelled in c) and d).}
     \label{fig:onera_ridges}
\end{figure}

In this case, the behaviour of the ridge functions as we extrapolate away from the training boundary conditions is important. To explore this, polynomial ridge functions are trained on the $M_{train}=36$ point DoE shown in the inset of Figure~\ref{fig:onera_viz}. The transonic nature of these flows results in a noticeable step change in the flow behaviour as the freestream Mach number increases, and at least fifth degree polynomials are found to be necessary to properly capture the surface Mach number's response. The two dimensional ridge functions are used to estimate the Mach number on the wing surface at the $s_3/S=0.2$ span-wise slice highlighted in Figure~\ref{fig:onera_viz}, with the estimates for two different DoE points from the \textit{test} set shown in Figures~\ref{subfig:onera_profle_interp} and~\ref{subfig:onera_profile_extrap}. The corresponding two dimensional ridges for the locations labelled A and B are then visualised in Figures~\ref{subfig:onera_ridge_A} and~\ref{subfig:onera_ridge_B}. From these visualisations it is clear that DoE point 1 lies within the training distribution, hence the it's predictions (Fig.~\ref{subfig:onera_profle_interp}) can be considered to be interpolations. In this case, the predicted Mach number distribution agrees well with the true distribution.

Moving on to DoE point 2, here the ridge approximations must extrapolate away from the flow conditions seen during training. Unsurprisingly, the predicted distribution in Figure~\ref{subfig:onera_profile_extrap} agrees less well with the true distribution in this case, especially on the aft portion of the wing. This highlights a note of caution regarding polynomial ridge functions; as with many function approximation techniques, caution must be taken when using the underlying polynomials to extrapolate beyond the training data. If such a task is routinely necessary, it is possible to replace the polynomials with other classes of models with more stable behaviour under extrapolation, such as piecewise polynomials (splines) or kernel regressors. However, we shall leave this as a topic of future work. 

\subsection{Three-dimensional flowfield predictions}

\begin{figure}[ht]
	\centering
     \includegraphics[width=\linewidth]{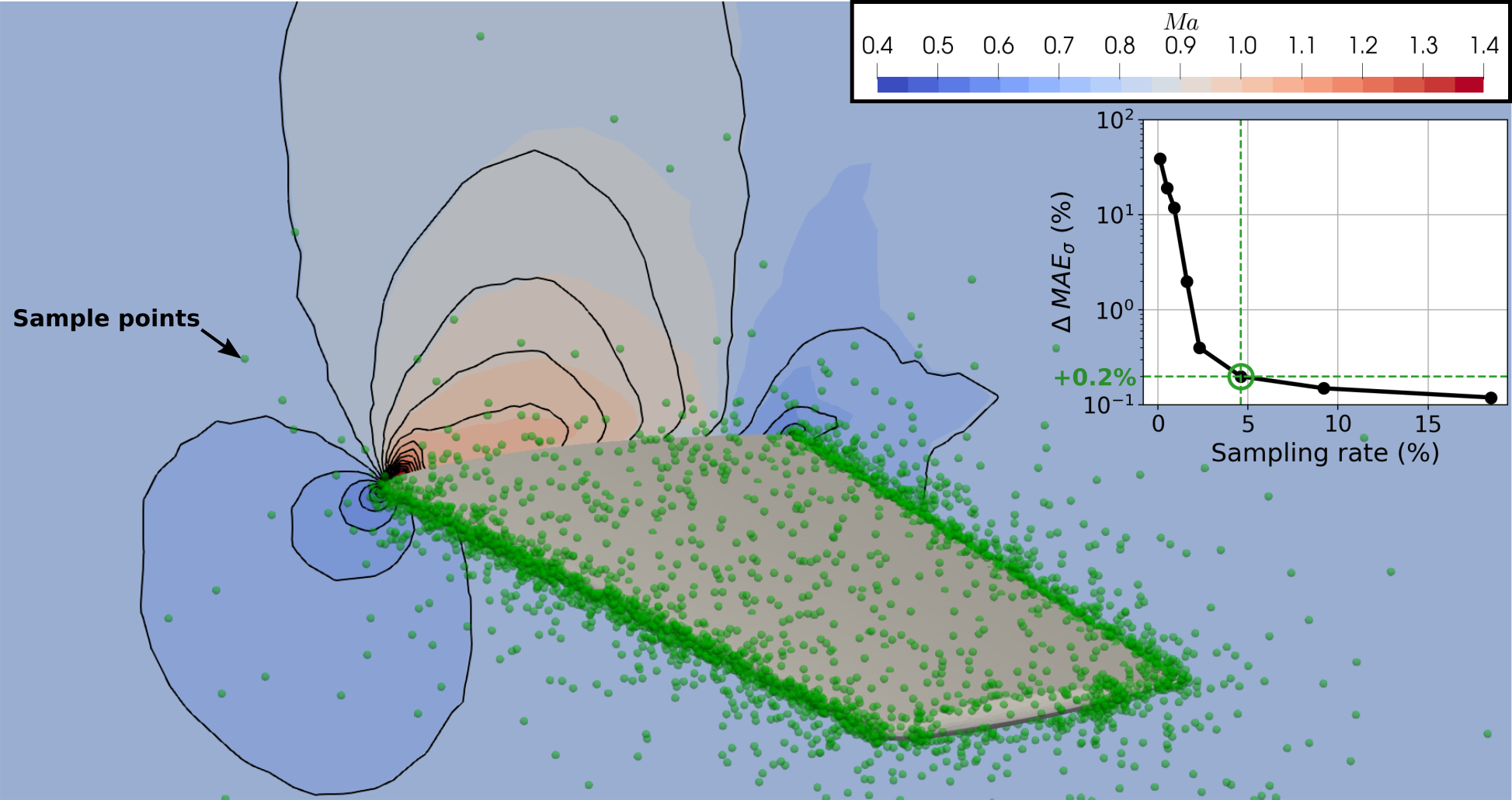} 
	\caption{A span-wise slice ($s_3/S=0.2$) of the upsampled transonic wing flowfield estimates of Mach number, for a DoE point from the \textit{test} set, with $Ma_\infty=0.77$ and $\alpha_\infty=4.6^{\circ}$. The $J=5000$ randomly subsampled points are shown in green. The inset figure shows the difference in normalised MAE between the subsampled-upsampled estimates, and the estimates with no subsampling ($J=N$), for different sampling rates.}
	\label{fig:onera_contours}
\end{figure}

When extending the framework to three dimensional flowfields, computational efficiency is crucial. To achieve a surrogate model for the full flowfield the subsampling and upsampling strategy is used once again. Ridge functions are fitted at the $J=5000$ randomly sampled points (a sampling rate of 4.6\%), with a covariance matrix used to upsample the ridge approximations to full three dimensional flowfield estimates. A slice of a flowfield estimate for a \textit{test} DoE point is shown in Figure~\ref{fig:onera_contours}. Comparing the colour contours and iso-lines, the prediction appears to be reasonable, with only a slight disagreement on the aft section of the suction surface.

The above result suggests the subsampling strategy is effective for three dimensional flows in addition to two dimensional ones. The inset figure within Figure~\ref{fig:onera_contours} shows the effect of sampling rate on the prediction error. For the present case, sampling rates of lower than $\approx 4\%$ lead to significantly increased errors. Nevertheless, sampling rates of 4-10\% still allow for significant computational cost savings. 
\section{Conclusions}

This paper exposes the idea of using spatially correlated ridge functions to rapidly estimate flowfields. The resulting data-driven framework could be trained on existing CFD data where available, or it could be integrated within a wider design of experiment, with new CFD data generated specifically for training. Once trained, the framework provides rapid flowfield predictions, which can be used for design space exploration, design optimisation tasks, performance predictions, or even to guide CFD mesh design. 

On the subsonic aerofoil test case, comparisons with the state-of-the-art convolutional neural network (CNN) suggest that the ridge function framework is able to achieve competitive predictive accuracy on unseen aerofoil designs. In addition to serving as a surrogate model, the learned ridge functions can aid understanding of the existing training data. \emph{Sufficient summary plots} can be viewed for any given point. The reduced dimensional nature of these plots lends itself to easy visualisation, allowing for easy comparison between designs, and new physical insights. Furthermore, the local ridge functions can be integrated to obtain new ridge functions for integral quantities such as a loss or drag coefficient. This allows new interpretable surrogate models for integral quantities to be generated on demand, without the need for further pre-processing of the CFD data. Such capability is advantageous in scenarios where the long-term storage of the full CFD dataset is problematic. The ridge functions can also provide \textit{insensitivity} information, informing us how a design can be altered without influencing the flow at a given location. 

The independent nature of each ridge function means that their training can be viewed as an embarrassingly parallel task. This makes the ridge function framework trivial to implement in a parallel fashion, allowing for excellent scaling with problem size. However, for larger problems, computing ridge functions for every single grid point is undesirable. Instead, ridge functions are obtained at a much smaller number of randomly subsampled points. The flow physics encoded within covariance matrices, computed from the training data, can be used to upsample the ridge functions' predictions back to the rest of the flowfield. The computation of the covariance matrices, and the subsequent use of them, involves a variety of linear algebra operations which can be implemented efficiently. For the three dimensional transonic wing test case, sampling rates around 5\% are achieved with minimal additional errors, allowing for significant reductions in training time.

The transonic wing test case demonstrates how ridge functions can be used to estimate flowfields with varying boundary conditions. In such cases, there is a danger of extrapolating too far from the training data, and the underlying polynomials' predictions can not be trusted here. If extrapolation is important, there is scope to replace the polynomials used in this work with alternative models. In addition to exploring this further,  a worthwhile area of future work would be to combine the approaches used in the two test cases, incorporating geometric and boundary condition input parameters together. 

\section*{Acknowledgments}
This research was supported in part through computational resources provided by The Alan Turing Institute and with the help of a generous gift from Microsoft Corporation. The authors were supported by The UKRI Strategic Priorities Fund under the EPSRC Grant EP/T001569/1, particularly the ``Digital twins for complex systems engineering'' theme within that grant and The Alan Turing Institute, and by the Lloyd’s Register Foundation-Alan Turing Institute programme on Data-Centric Engineering under the LRF grant G0095.

\appendix
\section{Convolutional Neural Network Setup}
\label{appdx:CNN}

This appendix provides more details on the convolutional neural network setup in Figure~\ref{fig:CNN_schematic}. This network is a modified form of that proposed by \citet{Thuerey2020}, which is based upon the U-Net architecture first proposed by \citet{Ronneberger2015}. Since the boundary conditions are scalar values, it might appear wasteful to repeat the values over the entireity of the $128\times 128 \times 3$ boundary condition input channels, and some CNN flowfield prediction frameworks such as those of \citet{Bhatnagar2019} inject the scalar values into the feature vector instead. However, \citet{Thuerey2020} claim that, while the network would eventually propagate the boundary information via the convolutional bow tie structure, specifying the redundant boundary condition information with skip connections allows for a more efficient training process.

The general design features of the convolutional blocks from \citet{Thuerey2020} are retained; in the encoder leaky ReLU activations functions are chosen to avoid the \textit{dying ReLU problem}\cite{Lu2020}, whilst standard ReLU functions are selected for the decoder, both with a slope of 0.2. Convolutional filter kernels of various sizes and strides are then used for encoding and decoding (See Fig.~\ref{fig:CNN_schematic}), with nearest neighbour upsampling followed by a regular convolution for the decoder blocks. To help mitigate overfitting, batch normalisation is used in all the blocks except for the first and last ones (1c and 1d), in addition to having a slight dropout rate of 0.01 for all layers. 

\subsection{Learning process}

To obtain the the weights (and biases) of the network, the learning process uses the \textit{Adam optimiser}\cite{Kingma2014} to minimise the Huber loss function given in \eqref{eqn:CNN_loss}, with the loss averaged over all $M_{train}$ number of training designs and $N=128^2$ grid points. A learning rate of $4\times10^{-4}$ is used, which is then reduced by a factor of ten after 500 epochs, and a batch size of ten is chosen. 

\begin{figure}[ht]
	\centering
     \includegraphics[width=0.6\linewidth]{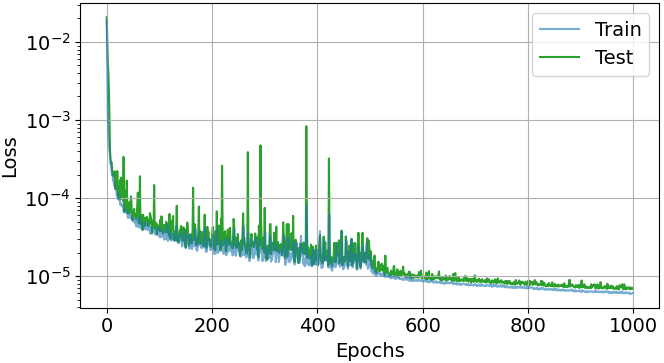}
     \caption{Convergence history for the 488k parameter CNN trained on the $\alpha_\infty=10^{\circ}$ dataset, with $M_{train}=500$.}
     \label{fig:CNN_convergence}
\end{figure} 

The learning curve for the 488k parameter network is plotted in Figure~\ref{fig:CNN_convergence}; the convergence behaviour appears to be satisfactory, and the small difference between train and test loss suggests the network isn't suffering from excessive overfitting. Similar behaviour is observed for the other network sizes explored in this paper.

\subsection{Effect of network size}

To allow for a fair comparison between the ridge function framework and the CNN, the performance for various size/complexity networks is examined in order to find an optimal size. The network's size is altered by changing the number of channels, with the number of channels in the $i^{th}$ layer given by $C_i=2^ec_i$. The constants $c_i$ are defined in Table~\ref{tab:CNN_size}, and the network's size is varied by adjusting the exponent $e$, with values of $e=\{2,3,4,5\}$ tested. The variation in performace

\begin{table}[ht]
\centering
\caption{Multiplier $c$ used to set the number of channels in each layer of the convolutional neural network. The number of channels in the $i^{th}$ layer is given by $C_i=2^ec_i$, where the exponent $e$ is a scalar parameter used to set the model complexity.} \label{tab:CNN_size}
\begin{tabular}{ccccccccc}
\toprule
\multirow{2}{*}{Encoder} & Layer & 1c & 2c & 3c & 4c & 5c & 6c & 7c  \\
\cline{2-9} 
& $c_i$ & 1 & 2 & 2 & 4 & 8 & 8 & 8 \\
\midrule
\multirow{2}{*}{Decoder} & Layer & 1d & 2d & 3d & 4d & 5d & 6d & 7d  \\
\cline{2-9} 
& $c_i$ & 8 & 16 & 16 & 8 & 4& 4& 4 \\
\bottomrule
\end{tabular}
\end{table}

\begin{figure}[ht]
	\centering
    	\includegraphics[width=0.4\linewidth]{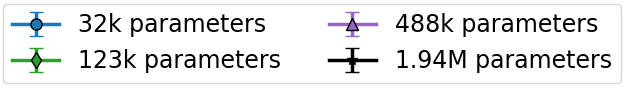} \vspace{-8pt} \\
 	\subfloat[MAE of $C_p$ predictions
  		\label{subfig:CNN_Cp_mae}]{%
  		\includegraphics[width=0.325\linewidth]{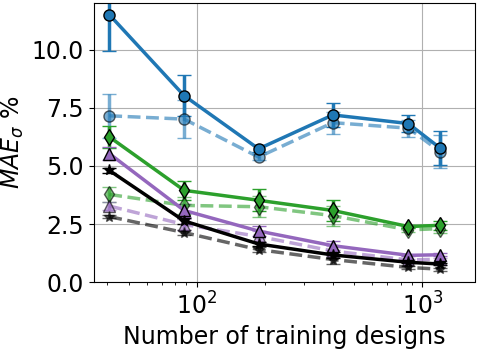}}   \hfil     	
 	\subfloat[MAE of $u/U_\infty$ predictions
  		\label{subfig:CNN_u_mae}]{%
  		\includegraphics[width=0.325\linewidth]{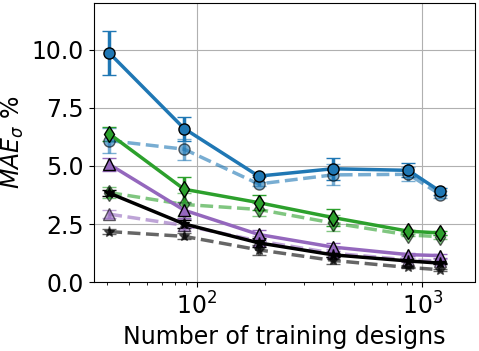}} \\
 	\subfloat[MAE of $\nu_t/\nu_\infty$ predictions
  		\label{subfig:CNN_nut_mae}]{%
  		\includegraphics[width=0.325\linewidth]{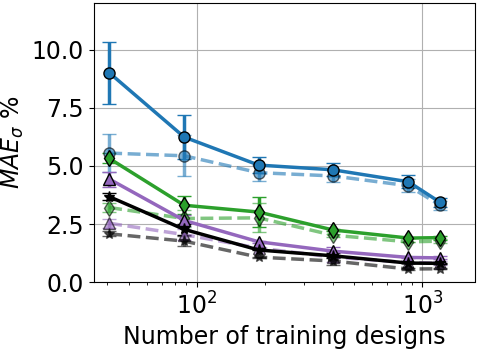}} \hfil
 	\subfloat[Training time
  		\label{subfig:CNN_time}]{%
  		\includegraphics[width=0.325\linewidth]{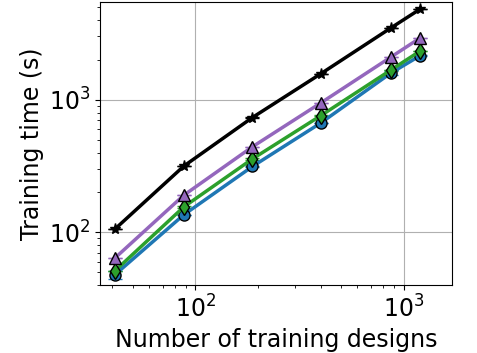}}     	    	
	\caption{Prediction errors and training time for different size convolutional neural networks, for the $\alpha_\infty=10^{\circ}$ dataset. Solid lines indicate test errors, and dashed lines training errors. Results are averaged over three train/test folds, and the error bars denote the standard deviation in $MAE_\sigma$ across  these folds. The size of the network is altered by adjusting the number of channels via the exponent $e$.}
	\label{fig:CNN_peform}
\end{figure}

The normalised MAE, defined in \eqref{eqn:MAE}, is measured on the training and test sets for the four different network sizes, with the results plotted in Figure~\ref{fig:CNN_peform}. Generally, the larger networks achieve lower predictive errors. However, this is at the cost of longer training times (Fig.~\ref{subfig:CNN_time}), with the 1.94M parameter network taking up to 80 minutes to train. The larger networks also exhibit more stable predictions, with less variance in their MAE scores across the three train/test folds. To achieve a reasonable compromise between accuracy and training cost, the 488k parameter is chosen as the benchmark for comparisons elsewhere in this paper. A similar pattern between MAE, training times, and the number of training designs is also observed. The chosen value of $M_{train}=500$ is seen to offer a good compromise between low MAE scores and short training time.

\section{Integrating Ridge Functions}
\label{appdx:integrate_ridge}

This appendix presents a procedure for identifying a dimension reducing subspace for the mass-averaged pressure loss coefficient $\bar{Y}_p$, defined in \eqref{eqn:Yp_int}. A more general treatment on the subspaces of an \textit{integral of a scalar field quantity} can be found in \citet{wong2020embedded}. For simplicity, we stick to a two dimensional domain, but the approach is readily generalisable to three dimensions. Recalling \eqref{eqn:Yp_int}, we wish to obtain the mass-averaged loss coefficient $Y_p$, at a given axial chord location $C_x$
\begin{equation} \label{eqn:Yp_int_appen}
\bar{Y}_p =  \frac{1}{\dot{m}}\int_{C_1} u(s_1,s_2) Y_p(s_1,s_2)ds_2, 
\end{equation}
where the mass flow rate across the plane, $\dot{m} = \int_{C_1} u(s_1,s_2) ds_2$, is taken to be a constant here\footnote{The standard deviation of $\dot{m}$ across the $C_1$ plane is less than 0.2\% of the mean.}. The local axial velocity and loss coefficient are approximated with two ridge approximations
\begin{equation} \label{eqn:u_Yp_approx}
\begin{split}
u(s_1,s_2) &\approx g_{s_1,s_2}\left(\mathbf{W}^T_{s_1,s_2} \mathbf{x} \right) \\
Y_p(s_1,s_2) &\approx h_{x,y}\left(\mathbf{U}^T_{s_1,s_2} \mathbf{x} \right).
\end{split}
\end{equation}
Here the input vector $\mathbf{x} \in \mathbb{R}^d$ has its usual meaning, whilst $\mathbf{W}^T \in \mathbb{R}^{n_w\times d}$ and $\mathbf{U}^T \in \mathbb{R}^{n_u\times d}$ denote the subspaces for axial velocity and loss coefficient respectively, where naturally $n_w \ll d$ and $n_u \ll d$. Substituting \eqref{eqn:u_Yp_approx} into \eqref{eqn:Yp_int_appen} yields
\begin{equation}
\bar{Y}_p = \rho_\infty \int_{C_1} g_{s_1,s_2}\left(\mathbf{W}^T_{s_1,s_2} \mathbf{x} \right) h_{s_1,s_2}\left(\mathbf{U}^T_{s_1,s_2} \mathbf{x} \right) ds_2.
\end{equation}

To identify the dimension reducing subspace for $\bar{Y}_p$ with respect to its constituent input parameters $\mathbf{x}$, we need to compute the \textit{averaged outer product of its gradient} \cite{constantine2015active} -- a covariance matrix given by
\begin{equation}
\mathbf{C} = \int_{\mathbb{R}^d} \left( \nabla_\mathbf{x} \bar{Y}_p \left(\mathbf{x}\right) \right) \left( \nabla_\mathbf{x} \bar{Y}_p \left(\mathbf{x}\right) \right)^T \omega(\mathbf{x}) d\mathbf{x},
\end{equation} 
where $\omega(\mathbf{x})$ denotes the distribution associated with the input parameters $\mathbf{x}$. Assuming the existence of a design of experiment with $M$ distinct CFD evaluations for different input parametrisations, and taking $\omega(\mathbf{x})$ to be a uniform distribution here, we can approximate $\mathbf{C}$ with its finite sample estimate
\begin{equation} \label{eqn:finite_C}
\mathbf{C} \approx \frac{1}{M} \sum_{m=1}^M \left( \nabla_\mathbf{x} \bar{Y}_p \left(\mathbf{x}_m\right) \right) \left( \nabla_\mathbf{x} \bar{Y}_p \left(\mathbf{x}_m\right) \right)^T.
\end{equation}
The gradient evaluations can be written as 
\begin{equation}
\begin{split}
\nabla_\mathbf{x} \bar{Y}_p \left(\mathbf{x}_m\right)   &= \rho_\infty \int_{\mathbb{R}^d} \pdv{\mathbf{x}_m} \left[ g_{s_1,s_2}\left(\mathbf{W}^T_{s_1,s_2} \mathbf{x}_m \right) h_{s_1,s_2}\left(\mathbf{U}^T_{s_1,s_2} \mathbf{x}_m \right)  \right] ds_2 \\
& \begin{aligned}
{}= \rho_\infty \int_{\mathbb{R}^d} & \left[ \mathbf{W}_{s_1,s_2}\nabla_{\mathbf{w}} g_{s_1,s_2}\left(\mathbf{w}_m \right) h_{s_1,s_2}\left(\mathbf{u}_m \right) + \mathbf{U}_{s_1,s_2}\nabla_{\mathbf{u}} h_{s_1,s_2}\left(\mathbf{u}_m \right) g_{s_1,s_2}\left(\mathbf{w}_m \right) \right]  ds_2,
\end{aligned}
\end{split}
\end{equation}
where $\mathbf{w}_m=\mathbf{W}^T \mathbf{x}_m$ and $\mathbf{u}_m=\mathbf{U}^T \mathbf{x}_m$ are the reduced dimensional co-ordinates for the $m^{th}$ design in the $Y_p(s_1,s_2)$ and $u(s_1,s_2)$ subspaces respectively. If we further assume that spatially, there are $N$ nodes along which this integration has to be performed, in which case we have the set
\begin{equation}
\left\lbrace g_i, h_i, \mathbf{W}_i, \mathbf{U}_i, \nabla_{\mathbf{w}}g_i, \nabla_{\mathbf{u}}h_i,  \right\rbrace_{i=1}^N,
\end{equation}
we approximate
\begin{equation}
\nabla_\mathbf{x} \bar{Y}_p \left(\mathbf{x}_m\right)  \approx \rho_\infty \sum_{i=1}^N \left[ \mathbf{W}_i \nabla_{\mathbf{w}} g_ih_i + \mathbf{U}_i \nabla_{\mathbf{u}} h_ig_i \right] \varphi_i
\end{equation}
where $\varphi_i$ denotes appropriately selected quadrature weights depending on the spatial location of the nodal centres. Plugging this into \eqref{eqn:finite_C} yields
\begin{equation}
\begin{split}
\mathbf{C} & \approx \mathbb{E} \left[ \left( \rho_\infty \sum_{i=1}^N \left[ \mathbf{W}_i \nabla_{\mathbf{w}} g_ih_i + \mathbf{U}_i \nabla_{\mathbf{u}} h_ig_i \right] \varphi_i \right) \left( \rho_\infty \sum_{k=1}^N \left[ \mathbf{W}_k \nabla_{\mathbf{w}} g_kh_k + \mathbf{U}_k \nabla_{\mathbf{u}} h_kg_k \right] \varphi_k \right)^T \right] \\	
&\begin{aligned}
{}={}&\rho_\infty^2 \sum_{i=1}^N\sum_{k=1}^N \varphi_i \varphi_k \mathbb{E}\left[  h_i h_k \mathbf{W}_i \nabla_{\mathbf{w}}g_i \nabla_{\mathbf{w}}g_k^T \mathbf{W}_k^T  \right] + 
\rho_\infty^2 \sum_{i=1}^N\sum_{k=1}^N \varphi_i \varphi_k \mathbb{E}\left[ h_i g_k \mathbf{W}_i  \nabla_{\mathbf{w}}g_i \nabla_{\mathbf{u}}h_k^T \mathbf{U}_k^T  \right] + \\
&\rho_\infty^2 \sum_{i=1}^N\sum_{k=1}^N \varphi_i \varphi_k \mathbb{E}\left[ g_i h_k \mathbf{U}_i  \nabla_{\mathbf{u}}h_i \nabla_{\mathbf{w}}g_k^T \mathbf{W}_k^T \right] + 
\rho_\infty^2 \sum_{i=1}^N\sum_{k=1}^N \varphi_i \varphi_k \mathbb{E}\left[  g_i g_k \mathbf{U}_i \nabla_{\mathbf{u}}h_i \nabla_{\mathbf{u}}h_k^T \mathbf{U}_k^T \right]  
\end{aligned}
\end{split}
\end{equation}

This equation is easy to implement, as long as the gradients $\nabla_{\mathbf{w}} g$ and $\nabla_{\mathbf{u}} h$ can be computed. When applying this approach in Section~\ref{sub:physics}, $g$ and $h$ are taken to be the underlying polynomials of the ridge function approximations, rather than the posterior mean of the ridge kernels as is the case in the rest of the paper. This is done since the gradients of orthogonal polynomials are readily available. Finally, an eigendecomposition of $\mathbf{C}$ is performed, and the leading eigenvectors with the largest eigenvalues are chosen to form $\mathbf{W}_{int}$, the dimension reducing subspace for $\bar{Y}_p$. In the present case, the first eigenvalue was found to contribute 99.3\% of the trace. Therefore, a one dimensional dimension reducing subspace was deemed  to be sufficient.

\bibliography{library}

\end{document}